\documentclass[10pt, a4paper, twocolumn, aps, pra, showpacs,
longbibliography, superscriptaddress, nofootinbib]{revtex4-1}
\pdfoutput=1
\usepackage[utf8x]{inputenc}
\usepackage{microtype}
\usepackage{ucs}
\usepackage{amsmath}
\usepackage{amsfonts}
\usepackage{amssymb}
\usepackage{makeidx}
\usepackage{cellspace,booktabs,array}
\usepackage{lipsum}
\usepackage{bm}
\usepackage[colorlinks=true,urlcolor=blue,citecolor=blue]{hyperref}
\usepackage{graphicx}
\usepackage[dvipsnames]{xcolor}
\usepackage{mathtools}
\usepackage{subcaption}
\usepackage{bbm}

\usepackage[misc]{ifsym}
\allowdisplaybreaks 

\usepackage{siunitx}
\usepackage{tabularx}

\makeatletter
\def\@fnsymbol#1{\ensuremath{\ifcase#1\or \text{\Letter} \or \ddagger\or
   \mathsection\or \mathparagraph\or \|\or **\or \dagger\dagger
   \or \ddagger\ddagger \else\@ctrerr\fi}}
\makeatother

\begin{document}

\begin{abstract}
Artificial spiking neural networks have found applications in areas where the temporal nature of activation offers an advantage, such as time series prediction and signal processing.
To improve their efficiency, spiking architectures often run on custom-designed neuromorphic hardware, but, despite their attractive properties, these implementations have been limited to digital systems.
We describe an artificial quantum spiking neuron that relies on the dynamical evolution of two easy to implement Hamiltonians and subsequent local measurements.
The architecture allows exploiting complex amplitudes and back-action from measurements to influence the input.
This approach to learning protocols is advantageous in the case where the input and output of the system are both quantum states. We demonstrate this through the classification of Bell pairs which can be seen as a certification protocol.
Stacking the introduced elementary building blocks into larger networks combines the spatiotemporal features of a spiking neural network with the non-local quantum correlations across the graph.
\end{abstract}

\date{\today}
\author{Lasse Bjørn Kristensen}\thanks{email: lbk@phys.au.dk}
\affiliation{Department of Physics and Astronomy, Aarhus University, DK-8000 Aarhus C, Denmark}
\author{Matthias Degroote}
\affiliation{Department of Chemistry, University of Toronto, Toronto, Ontario M5G 1Z8, Canada}
\affiliation{Department of Computer Science, University of Toronto, Toronto, Ontario M5G 1Z8, Canada}
\affiliation{Department of Chemistry and Chemical Biology, Harvard University, Cambridge, MA 02138,USA}
\author{Peter Wittek}
\affiliation{Rotman School of Management, University of Toronto, Toronto, Ontario, Canada}
\affiliation{Creative Destruction Lab, Toronto, Ontario, Canada}
\affiliation{Vector Institute for Artificial Intelligence, Toronto, Ontario M5S 1M1, Canada}
\affiliation{Perimeter Institute for Theoretical Physics, Toronto, Ontario N2L 2Y5, Canada}
\author{Alán Aspuru-Guzik}
\affiliation{Department of Chemistry, University of Toronto, Toronto, Ontario M5G 1Z8, Canada}
\affiliation{Department of Computer Science, University of Toronto, Toronto, Ontario M5G 1Z8, Canada}
\affiliation{Department of Chemistry and Chemical Biology, Harvard University, Cambridge, MA 02138,USA}
\affiliation{Vector Institute for Artificial Intelligence, Toronto, Ontario M5S 1M1, Canada}
\affiliation{Zapata Computing Inc., Cambridge, MA 02139, USA}
\affiliation{Canadian Institute for Advanced Research, Toronto, Ontario M5G 1Z8, Canada}
\author{Nikolaj T. Zinner}
\affiliation{Department of Physics and Astronomy, Aarhus University, DK-8000 Aarhus C, Denmark}
\affiliation{Aarhus Institute of Advanced Study, Aarhus University, DK-8000 Aarhus C, Denmark}

\title{An Artificial Spiking Quantum Neuron}

\maketitle

\section*{Introduction}
As Moore's law slows down~\cite{waldrop2016chips}, increased attention has been put towards alternative models for solving computationally hard problems and analyzing the ever growing stream of data~\cite{gantz2012digital,hashem2015rise}. One significant example has been the reinvigoration of the field of machine learning: neuromorphic models, inspired by biology, found applications in a large host of fields~\cite{krizhevsky2012imagenet,sutskever2014sequence}. In parallel, quantum computing has been taking significant steps moving from a scientific curiosity towards a practical technology capable of solving real-world problems~\cite{preskill2018quantum}. Given the prominence of both fields, it is not surprising that a lot of work has gone into exploring their parallels, and how one may be used to enhance the other. One such synergy has emerged in the field of quantum machine learning~\cite{biamonte2017quantum, dunjko2018machine,kapoor2016quantum}. Recent results aim to mimic the parametric, teachable structure of a neural network with a sequence of gates on a set of qubits~\cite{schuld2018circuit,killoran2018continuous,tacchino2019artificial} or on a set of photonic modes~\cite{steinbrecher2019quantum}.
A subset of these algorithms focus on quintessentially quantum problems: the input to the learning model is a quantum state and so is its output. This scenario is relevant in building and scaling experimental devices and it is often referred to as quantum learning~\cite{monras2016inductive,albarran2018measurement}.\\

We take a slightly different approach to quantum learning wherein the structure of the network manifests itself as interactions between qubits in space rather than as gates in a circuit diagram. Specifically, we will present a small toolbox of simple spin models that can be combined into larger networks capable of neuromorphic quantum computation. To illustrate the power of such networks, a small example of such a `spiking quantum neural network' capable of comparing two Bell states is presented, a task which could have applications in both state preparation and quantum communication.
The term `spiking' refers to the temporal aspect in the functioning of the model during the activation of the neuron, akin to the classical spiking neural networks~\cite{maass1997networks}. As illustrated in the example, a fundamental property of these networks is that they generate entanglement between the inputs and outputs of the network, thus allowing measurement back-action from standard measurements on the output to influence the state of the input in highly non-trivial ways.
The proposed model for spiking quantum neural networks is amenable to implementation in a variety of physical systems, e.g., using superconducting qubits~\cite{kjaergaard2019superconducting, kounalakis2018tuneable,wallraff2007sideband}.\\

A key inspiration for the constructions presented in this paper is the advent of dedicated classical hardware for simulating spiking classical neural networks, including implementations from both Intel and IBM~\cite{Roy2019,Merolla2014,Davies2018,Benjamin2014}. These systems emulate the function of a biological spiking neural network through networks of small neuron-like computing resources. They can be roughly grouped into digital systems simulating the dynamics of spiking networks using binary variables and in discrete time-steps~\cite{Merolla2014,Davies2018}, and analog circuitry emulating spiking behaviour of physical observables in continuous time~\cite{Benjamin2014}. The main focus of this paper is to extend the concept of the discretized models to the quantum domain, facilitating the use of purpose-built neuromorphic systems for applications within quantum learning, and providing the models access to quantum ressources like entanglement. Additionally, an outlook towards the implementation of continuous-time computational quantum dynamics is also briefly discussed.\\

Note that several similar proposals for real-space quantum neural networks do exist. One prominent example is within the field of quantum memristors, where a spiking quantum neuron was recently constructed, and networks of these objects were proposed~\cite{gonzalez2019quantized}. The central distinction to our scheme is the degrees of freedom under consideration. Whereas our scheme is based on generic qubits, memristive schemes revolve around the dynamics of voltages and currents. This means that the operations of our neural networks below will be more easily interpretable in a quantum computing context in comparison to these dissipative memristive schemes. A more apt comparison may therefore be a recent proposal for implementing a quantum perceptron through adiabatic evolution of an Ising model~\cite{torrontegui2019unitary}. Indeed, the interaction implemented in that proposal very closely resembles the operation of the first spin-model presented below. However, the nature of the adiabatic protocol makes the path towards connecting such building blocks into a larger dynamical network unclear. Neither case fully captures the properties of the spiking quantum neural network proposed here.

\section*{Results}
\subsection*{Defining the Building Blocks}
The first step towards a neuromorphic quantum spin model is the construction of neuron-like building blocks. In other words, we need objects capable of sensing the state of a multi-spin input state and encoding information about relevant properties of this input into the state of an output spin. Additionally, we will require that this operation does not disturb the state of the input. This additional property is partially motivated by a similar property of the neurons used in e.g. classical feed forward networks, which similarly only exerts influence on the state of the network through their output. Furthermore, we show in Sec.~\ref{sec:Comparison_Network} below that the preservation of inputs allow for interesting and non-trivial effects of the entanglement between the generated output and the preserved input. The cost of this preservation is a set of restrictions on each building block, as described in more detail below. \\

Inspired by the way classical neurons activate based on a (weighted) sum of their inputs, the first building block will be one that flips the state of its output spin, depending on how many of the input spins are in the `active' $\left| \uparrow \right>$ -state. The second building block, on the other hand, measures relative phases of components in the computational (i.e. the $\sigma^z$) basis, and thus has no classical analogue.

\subsection{Neuron 1: Counting Excitations}
\label{sec:Exc_Neuron}
\begin{figure}[htbp]
  \centering
   \includegraphics[width=\columnwidth]{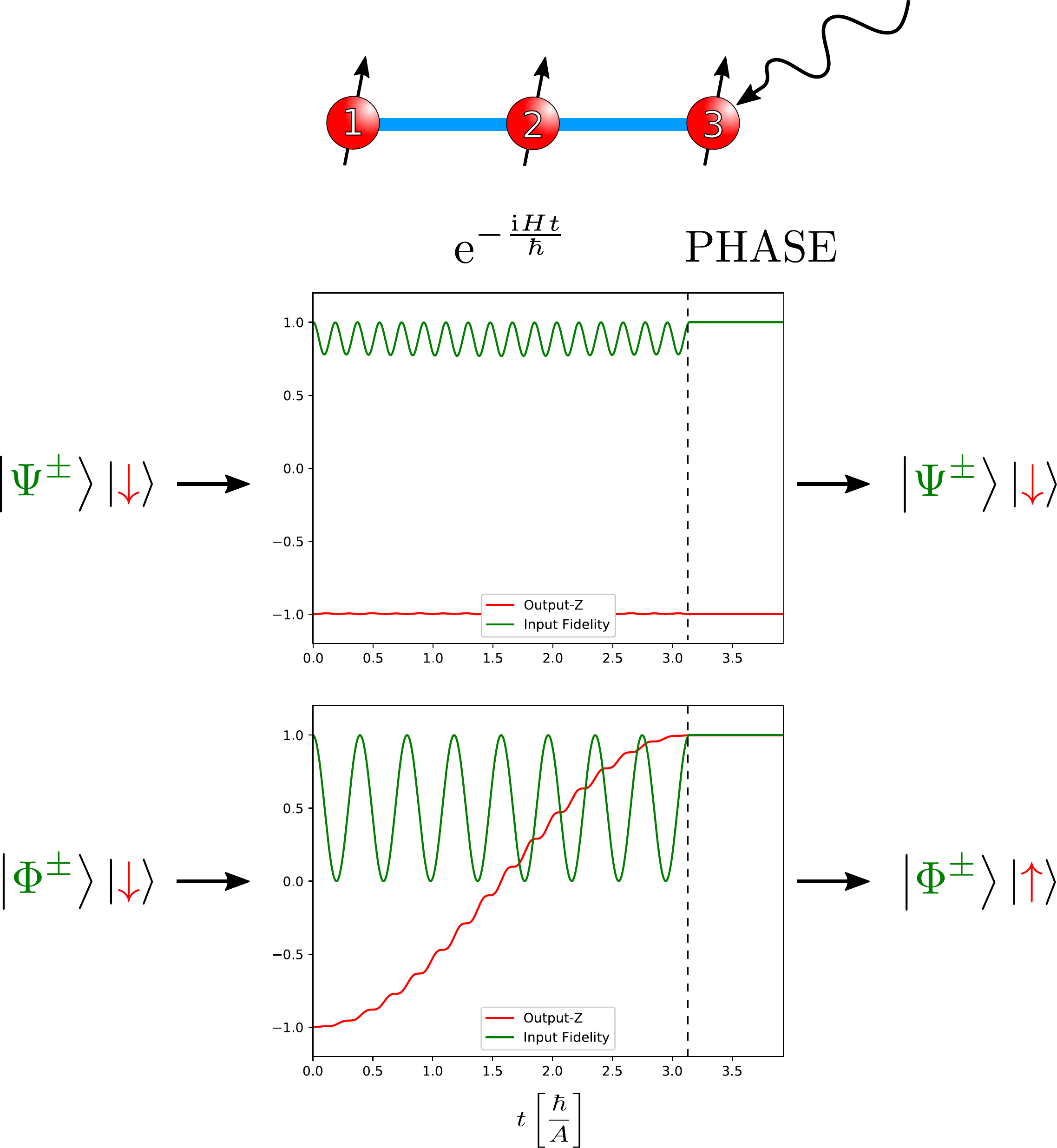}
   \caption{\textbf{Dynamics of Neuron 1.} Schematic depiction of the excitation-number detecting neuron (top) and plots of the time-evolution of the output-qubit state (red) and the overlap between the state of the input qubits and their initial state (green) for the four Bell states during the operation of the neuron. As illustrated in these plots, the state of the output is either flipped or not depending on whether the input contains an odd (middle) or even (bottom) number of excitations. In contrast, the input qubits return to their initial states in all four cases. Parameters used are $l=17$, $k=8$, which yields an average operational fidelity of $99.98\%$ in the absence of noise.}
   \label{fig:Exc_Neuron}
\end{figure}

In analogy to the thresholding behaviour of classical spiking neurons, we start by constructing a spin system that is capable of detecting the number of excitations (i.e. the number of inputs in the state $\left| \uparrow \right>$) and exciting its output spin conditional on this information. As shown in Supplementary Note~\ref{sec:Exc_Neur}, this behaviour can be implemented using dynamical evolution driven by the Hamiltonian
\begin{align}
H_{\text{Exc}} =&\, \frac{J}{2} \left( \sigma_1^x \sigma_2^x + \sigma_1^y \sigma_2^y + \sigma_1^z \sigma_2^z \right) + \beta \, \sigma_2^z \sigma_3^z \nonumber\\
&+ A \cos\left( \frac{2 \beta}{\hbar} t \right) \sigma_3^x \; ,
\label{eq:H_Exc}
\end{align}
assuming the evolution is run for a time $\tau=\pi \hbar / A$ and that the Hamiltonian fulfils some restrictions on the interaction strengths $J, \beta, A$ that will be discussed below. In this model, we label spins 1 and 2 as the input, and spin 3 as the output. The intuition behind this model is that the Heisenberg interaction between the inputs sets up energy differences among the four possible Bell state of the inputs. Through the $\sigma_2^z \sigma_3^z$ coupling to the output, these differences then influence the energy cost of flipping the output spin, resulting in the cosine drive on the output only being resonant when the input qubits are in certain states. The result is that the driving induces flips in the output qubit if and only if the input is in a Bell state with an even number of excitations. Since the conditionality is a resonance/off-resonance effect, the detuning of the undesired transitions needs to be much larger than the strength of the driving, which leads to the criterion
\begin{align}
\Delta_{\pm} = \left| 2\beta \pm 2 \sqrt{J^2 + \beta^2} \right| \gg A \; ,
\label{eq:Detuning_Restriction_Neuron_1}
\end{align}
which is naturally fulfilled whenever the driving-strength $A$ is much smaller than the chain interaction strength $J$.\\
Due to dynamical phases, the requirement that superpositions of input states should be preserved adds two additional constraints for the parameters of the model. Specifically, conservation of relative phases within the subspaces of inputs that either flips (``above threshold'') or does not flip (``below threshold'') the output yields
\begin{align}
\beta &= k A & & k \in \mathbb{Z}\nonumber\\
J &= \pm \sqrt{l^2-k^2} \, A & & l \in \mathbb{Z} \; .
\end{align}
If these constraints are fulfilled, the only non-trivial phase will be a coherent phase between the above-threshold and below-threshold subspaces of $-i (-1)^{k+l} \exp\left(-i Jt/\hbar\right)$, where $t$ is the time elapsed during detection. Since these two subspaces are now distinguished by the state of the output, correcting for this phase is just a matter of performing the corresponding phase-gate on the output qubit. In the special case where $\sqrt{l^2-k^2}$ is an integer, this reduces to performing a $\pi/2$-rotation of the output about the $z$-axis (see Supplementary Note~\ref{sec:Exc_Neur} for details).\\
When combined with this subsequent unconditional phase gate, the dynamical evolution induced by the Hamiltonian in \eqref{eq:H_Exc} is to coherently detect the parity of the number of excitations of the input, and to encode this information in the output spin, i.e. in conventional Bell-state notation:
\begin{align}
\left| \Psi^\pm \right> \left| \downarrow \right> \; &\longrightarrow \; \left| \Psi^\pm \right> \left| \downarrow \right> \nonumber\\
\left| \Phi^\pm \right> \left| \downarrow \right> \; &\longrightarrow \; \left| \Phi^\pm \right> \left| \uparrow \right> \; ,
\label{eq:Number_Detection}
\end{align}
where the first ket denotes the state of the inputs and the second ket the state of the output. The output is either fully excited or not excited at all by the evolution---hence we refer to this structure as a spiking quantum neuron, in analogy to similar objects from classical computing. Sample simulations showing the dynamics of the spiking process are shown in Fig.~\ref{fig:Exc_Neuron}. Note that the dynamics implementing this behaviour is linear unitary time evolution, thus the effect on general 2-qubit inputs follows from expressing the input in the Bell-state basis and applying the rules of \eqref{eq:Number_Detection}. For the neuron-parameters $l=17$ and $k=8$, the corresponding operation is performed with an average fidelity of $99.98\%$ when averaged over all possible 2-qubit inputs.
\subsection{Neuron 2: Detecting Phases}
\label{sec:Phase_Neuron}
While neuron 1 is a fully quantum mechanical object, capable of coherently treating superpositions in the inputs, the property that it detects---the number of excitations in the input---would be similarly well-defined for a classical neuron participating in a classical digital computation. However, the state of the two input qubits will also be characterized by properties that have no classical analogue, such as the relative phases of terms in a superposition state. The goal of the second neuron is to be able to detect these relative phases of states in the computational basis. Specifically, it aims to distinguish the states $\{ \left| \Psi^+ \right>, \left| \Phi^+ \right> \}$ from the states $\{ \left| \Psi^- \right>, \left| \Phi^- \right> \}$. Combining this detection-capability with the capabilities of the excitation-counting neuron of the previous section (exemplified by \eqref{eq:Number_Detection}) allows complete discrimination between the four Bell states.\\

\begin{figure}[htbp]
  \centering
   \includegraphics[width=0.95\columnwidth]{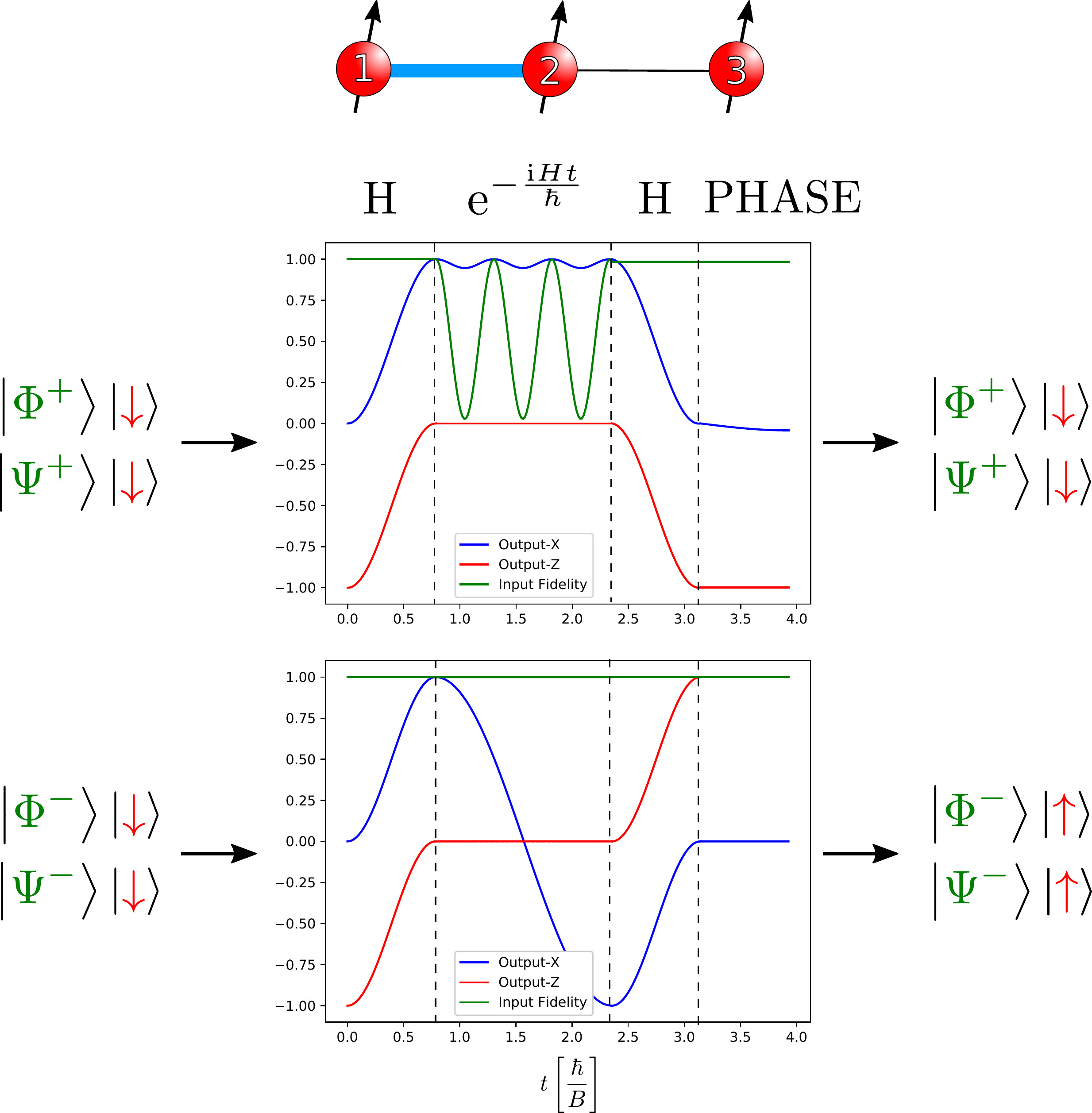}
   \caption{\textbf{Dynamics of Neuron 2.} Schematic depiction of the phase-detecting neuron (top) and plots of the time-evolution of the $X$- (blue) and $Z$-components of the output-qubit state as well as the overlap between the state of the input qubits and their initial state (green) for the four Bell states during the operation of the neuron. As illustrated in these plots, the state of the output is either flipped or not depending on whether the input contains a positive (top) or negative (bottom) relative phase. In contrast, the input qubits return to their initial states in all four cases. Parameters used are $n=82$, $m=3$, which yields an average operational fidelity of $99.07\%$ in the absence of noise. As detailed in Supplementary Note~\ref{sec:Phase_Neur}, small adjustments around these values can yield a slightly higher fidelity, in this case $99.58\%$.}
   \label{fig:Phase_Neuron}
\end{figure}

The operational principle of the phase-detection neuron is similar to that of the excitation-detection neuron: it relies on a combination of single-qubit gates and the unitary time-evolution generated by a Hamiltonian of the form:
\begin{align}
H_{\text{phase}} =& \,  \frac{J}{2} \left( \sigma_1^x \sigma_2^x + \sigma_1^y \sigma_2^y + \sigma_1^z \sigma_2^z \right) + \delta \, \sigma_2^x \sigma_3^x + B \, \sigma_3^z \; .
\label{eq:H_Phase}
\end{align}
As shown in Supplementary Note~\ref{sec:Phase_Neur}, running the dynamics of this Hamiltonian for a time $\tau=\pi \hbar/2B$ performs a $\left(-i Z\right)$-gate on the output qubit if and only if the state of input qubits are in the subspace spanned by $\left| \Psi^- \right>$ and $\left| \Phi^- \right> $. Thus by conjugating this operation with Hadamard gates on the output qubit and correcting for the $-i$-phase using a phase-gate (see Figure \ref{fig:Phase_Neuron}) yields the desired phase-detection operation:
\begin{align}
\left| \Psi^+ \right> \left| \downarrow \right> \; &\longrightarrow \; \left| \Psi^+ \right> \left| \downarrow \right> \nonumber\\
\left| \Phi^+ \right> \left| \downarrow \right> \; &\longrightarrow \; \left| \Phi^+ \right> \left| \downarrow \right>\nonumber\\
\left| \Psi^- \right> \left| \downarrow \right> \; &\longrightarrow \; \left| \Psi^- \right> \left| \uparrow \right>\nonumber\\
\left| \Phi^- \right> \left| \downarrow \right> \; &\longrightarrow \; \left| \Phi^- \right> \left| \uparrow \right> \; .
\label{eq:Phase_Detection}
\end{align}

The fundamental principle of operation is identical to the one of the excitation-counting neuron, in that the Hamiltonian once again contains three terms: a Heisenberg interaction to set up an energy spectrum that distinguishes the Bell states, an interaction that tunes the energy of the output qubit (i.e. qubit 3) dependent on the state of the inputs, and a single-qubit operator attempting to change the state of the output and succeeding if and only if the driving related to this term matches the energy cost of flipping the output. The only difference is that the interaction term now sets up an energy splitting between the spin states $\left| \pm \right> = \frac{1}{\sqrt{2}} \left( \left| \downarrow \right> \pm \left| \uparrow \right>\right)$ rather than the states $\left| \uparrow\right>/\left| \downarrow\right>$, hence the need for Hadamard gates to convert between the two bases. Note that all of these operations are again linear, thus \eqref{eq:Phase_Detection} specifies the operation also for general 2-qubit inputs. For the neuron-parameters $n=82$ and $m=3$, the corresponding phase-detection operation is performed with an average fidelity of $99.07\%$ when averaged over all possible 2-qubit inputs, with higher values achievable through small adjustments (see Supplementary Note~\ref{sec:Phase_Neur} for details).\\

As the operation of the phase-detection neuron also relies on resonance/off-resonance effects, a restriction of similar to \eqref{eq:Detuning_Restriction_Neuron_1} is present. Specifically, we require that
\begin{align}
\Delta = 2 \delta \gg B \; .
\end{align}
Additionally, the requirement that the state of the inputs should not be distorted by the operation of the neuron yields the requirement that the ratios between $J,\delta$ and $B$ should fulfil:
\begin{align}
J &= 2n & & n \in \mathbb{Z} \nonumber\\
\delta &= 2m & & m \in \mathbb{Z} \; .
\end{align}
with $n\gg m$. 
\subsection{State Comparison Network}
\label{sec:Comparison_Network}
Having introduced a set of computational building blocks above, we now aim to illustrate how these can be combined into larger networks in order to solve computation- and classification-problems. Specifically, we will illustrate how a network of these objects allows one to compare pairs of Bell states to determine if they are the same Bell state. As detailed below, such a network could play a central role in the certification of Bell-pair sources and quantum channels, and may also have potential applications for machine learning and state preparation tasks.\\
\begin{figure}[htbp]
  \centering
   \includegraphics[width=0.8\columnwidth]{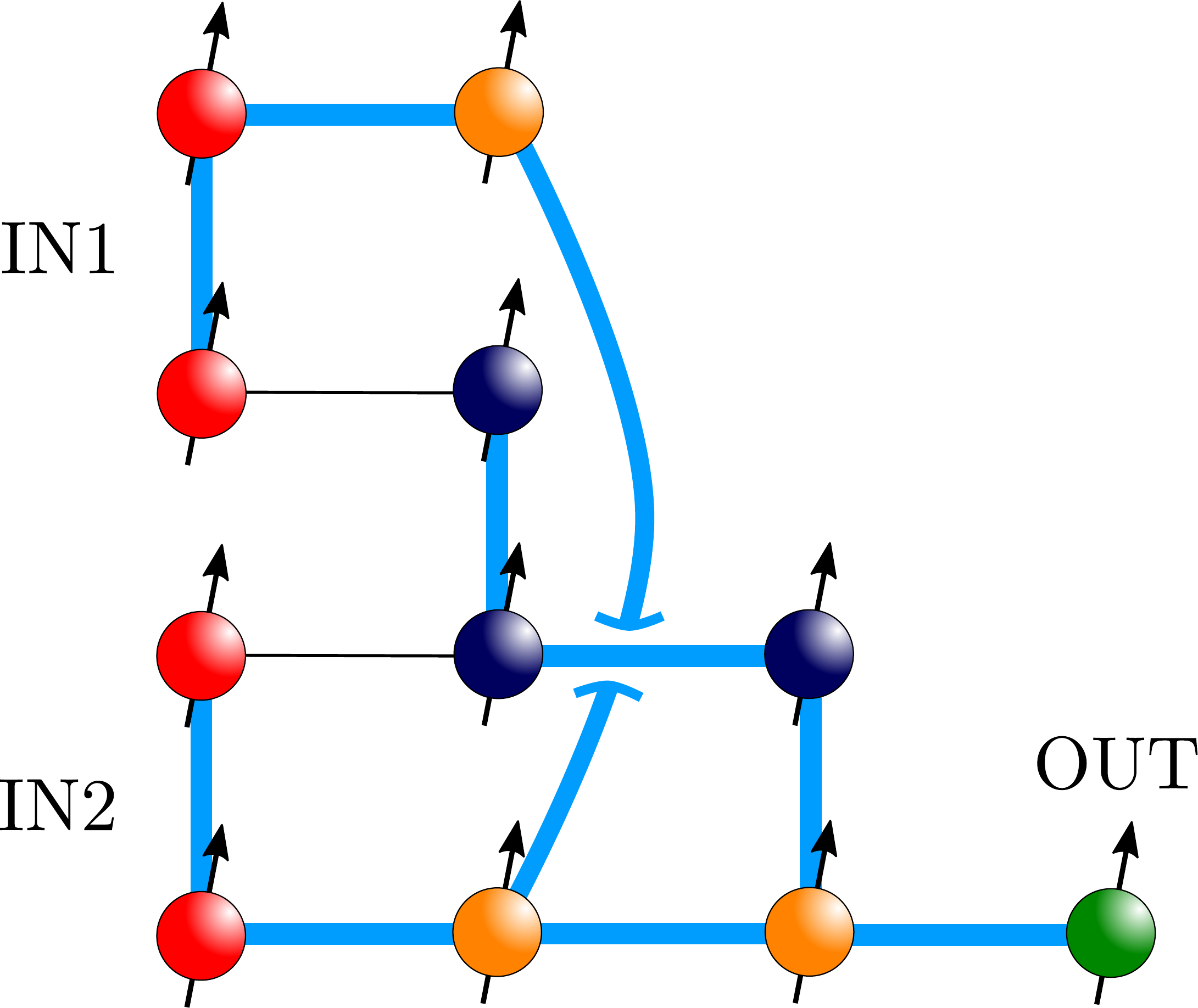}
   \caption{\textbf{Bell-state comparison network.} Schematic depiction of network for comparing pairs of Bell states. The left-most layer marked in red constitute the inputs to the network, with subsequent layers extracting and comparing information about either the phase (blue) or the excitation parity (orange) of the input states, with the result of the comparison stored in the green output qubit.}
   \label{fig:Comparator}
\end{figure}
The basic structure of the proposed network is depicted in Fig. \ref{fig:Comparator}, and consists of three layers. The first layer constitutes the input to the system. It is in these four qubits that the two Bell states to be compared are stored. Each pair is used as an input to both of the types of neurons detailed in sections \ref{sec:Exc_Neuron} and \ref{sec:Phase_Neuron}, with the output stored in two pairs of qubits in the second layer. In this way, sequentially running each of the two neuron operations extracts both the excitation-number parity and the relative phase of superpositions in the inputs and encodes it into the second layer. In other words, this layer ends up containing exactly the information needed to distinguish among the four Bell states. State comparison therefore boils down to detecting if the information extracted from one input matches that extracted from the other input. Since detecting if two qubits are in the same state (i.e. both $\left|\downarrow\right>$ or both $\left|\uparrow\right>$) boils down to checking the number of excitations modulo two, this comparison can be done using the neuron of Section \ref{sec:Exc_Neuron}. The third layer thus encodes two bits of information: whether the excitation parity of the two inputs match, and whether the relative phases match. Detecting if the two inputs were the same Bell state is thus a matter of detecting if both of these bits are in the $\left| \uparrow \right>$-state. This can be done using methods similar to those of Section \ref{sec:Exc_Neuron} -- see Supplementary Note~\ref{subsec:Final_layer} for details.\\

The result of the manipulations is that the output is put into the $\left|\uparrow\right>$-state if the two Bell-state inputs were identical and $\left| \downarrow \right>$ otherwise. Since all of the operations are achieved through linear, unitary dynamics, the behaviour for superposition inputs follow from this rule and linearity. This also implies that the network cannot compare arbitrary states, as is indeed prohibited by the no-cloning theorem of quantum mechanics. For instance, two identical inputs in a superposition in the Bell basis may return either $\left|\uparrow\right>$ or $\left|\downarrow\right>$ as output:
\begin{align}
\frac{1}{2} \left( \left| \Psi^+ \right> + \left| \Phi^- \right> \right)_{\text{IN1}} \left( \left| \Psi^+ \right> + \left| \Phi^- \right> \right)_{\text{IN2}} \left| \downarrow \right>_{\text{OUT}} \nonumber \\
 \; \longrightarrow  \frac{1}{2} \left( \left| \Psi^+ \right>_{\text{IN1}}\left| \Phi^- \right>_{\text{IN2}} + \left| \Phi^- \right>_{\text{IN1}}\left| \Psi^+ \right>_{\text{IN2}} \right) \left| \downarrow \right>_{\text{OUT}}\nonumber \\
 + \frac{1}{2} \left( \left| \Phi^- \right>_{\text{IN1}}\left| \Phi^- \right>_{\text{IN2}} + \left| \Psi^+ \right>_{\text{IN1}}\left| \Psi^+ \right>_{\text{IN2}} \right) \left| \uparrow \right>_{\text{OUT}} \; . \label{eq:Backaction}
\end{align}
This illustrates an interesting property of quantum neural networks, namely that entanglement between inputs and outputs means that measurements of the outputs may have dramatic effects on the state of the inputs. In this case, a measurement of $\left| \uparrow \right>$ in the output will always project the input qubits into the states corresponding to this output, i.e to identical Bell-state pairs. In this sense, a classifying quantum network like the one above will simultaneously be a projector onto the spaces corresponding to the states it is build to classify---a property that might prove helpful in, for instance, state preparation schemes. Note that this simple interpretation of measurement back-action follows from the fact that no other perturbations on the input state have been performed by the network during the computation, and hence from the corresponding non-disturbance requirement applied to each of the neurons. \\

A more concrete application of the network above is the certification of quantum channels and Bell-state sources. The ability to determine the reliability of resources such as Bell-state sources and quantum channels would be a practical benefit in many quantum communication and quantum cryptography applications. This is an active area of research: for instance, device-independent self-testing through Bell inequalities works for certain multipartite entangled states~\cite{supic2017simple}, or quantum template matching for the case where we have two possible template states~\cite{sasaki2001quantum,sentis2012quantumlearning}. Since the device presented above allows for the comparison of unknown systems with known-good ones, it is ideally suited to this kind of certification task.\\

Finally, it is worth noting that the comparative nature of the network means that the output of the network defines a kernel between 2-qubit quantum states. Specifically, given two inputs represented by amplitudes $\{ a_i \}$ and $\{b_i\}$ in the Bell-state basis, the probability of measuring $\left|\uparrow \right>$ in the output will be given by
\begin{align}
P_{\left| \uparrow \right>} = \sum_i \left| a_i \right|^2 \left| b_i \right|^2
\end{align}
which bears a strong resemblance to the classical ``expected likelihood'' kernel~\cite{jebara2004probability}. Considering this, comparison networks like the one above may also find applications within kernel-based quantum machine learning approaches~\cite{schuld2019quantum,havlicek2019supervised}.\\

\begin{figure}[htbp]
  \centering
   \includegraphics[width=0.65\columnwidth]{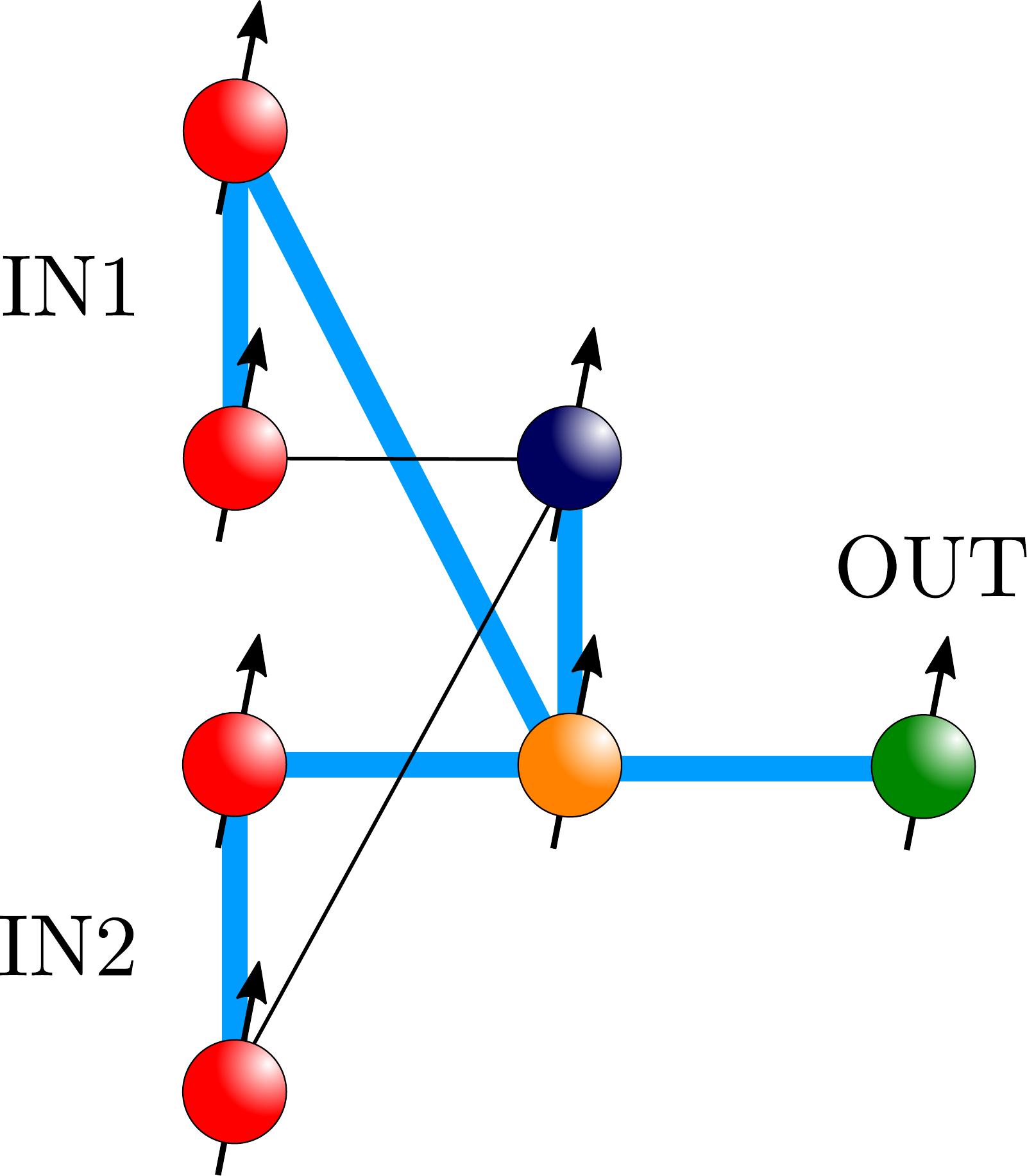}
   \caption{\textbf{Reduced Bell-state comparison network.} Schematic depiction of a network for comparing pairs of Bell states employing fewer qubits to do so than the one presented in the main text. The left-most layer marked in red constitute the inputs to the network, with subsequent layers extracting and comparing information about either the phase (blue) or the excitation parity (orange) of the input states, with the result of the comparison stored in the green output qubit.}
   \label{fig:Reduced_Comparator}
\end{figure}

One thing to note is that the design of the structure in Fig. \ref{fig:Comparator} is motivated by a desire for a one-step forward propagation of information between the layers of qubits, in analogy to the propagation of information in artificial classical neural networks. If each layer is allowed to probe the preceding layer more than once, a significant reduction in qubit overhead is possible. An example of this is the network depicted on Fig.~\ref{fig:Reduced_Comparator}. This network performs the same operation as the network in Fig.~\ref{fig:Comparator} while omitting the entire second layer. This is achieved by using the the same qubit as target for both of the neurons detecting a given property. In this way, a shared property between the two input pairs means the two sequential detections result in an even number of flips to the corresponding qubit of the middle layer -- either 0 or 2. On the other hand, non-identical properties will instead lead to precisely one flip. Thus the initial state of a qubit in the middle layer is preserved if and only if the property that it detects is identical between the two input pairs. The state of the output qubit is then determined by performing a flip if and only if both qubits in the middle layer have remained in their initial $\left| \downarrow  \right>$-state. In practice, this can be done using a similar generalization to the final step as the one used in the larger network -- See Supplementary Note~\ref{subsec:Final_layer} for details. Thus allowing multiple sequential probings of the inputs by the middle layer allows the qubit-count to be reduced by 4, though at the expense of a slight increase in the complexity of the protocol for forward propagation of information as well as an increase in the maximum number of connections required by a qubit from 3 to 4. 
\section*{Discussion}
We have presented a set of building blocks that detects properties of two-qubit inputs and encodes these properties in a binary and coherent way into the state of an output qubit. To illustrate the power of such spiking quantum neurons, we have presented a network of these building blocks capable of identifying if two Bell states are identical or not, and argued the usefulness of such comparison networks for quantum certification tasks within quantum communication and quantum cryptography. Additionally, we have seen how the entanglement of the inputs and output results in highly non-trivial effects on the inputs when a measurement is performed on the output of the network.\\

From the considerations above, several interesting questions arise. A main question might be how to scale up structures made from these and similar building blocks into larger networks capable of performing more complex quantum processing tasks. Concerning scaling, it seems reasonable to expect that the kind of intuitive reasoning behind the operation of the network presented here will become ever more challenging. As a result, it might be fruitful to take inspiration from the field of classical neural networks and design quantum networks whose operation depend on parameters. In this way, one can then adjust these parameters to make the network perform a certain task, in a way analogous to how both classical and quantum neural networks are trained. Since the Hamiltonians responsible for the operation of the neurons already contain a number of parameters, the architecture presented in this paper seems well-suited to such an approach. An interesting avenue of further research is therefore the generalization of the dynamical models presented here to tunable models capable of detecting other structures in multi-qubit inputs than the two-qubit Bell-state properties detected here, or to models capable of solving other relevant quantum-computing problems, for instance parity detection for error correction protocols. \\

A central challenge in such a learning-based approach would be how best to optimize the parameters of the model, and how to identify the class of operations that can be implemented by a given model. Much work is currently being dedicated to these questions in the context of variational algorithms in gate-based quantum computation~\cite{Peruzzo2014,Mcclean2016,schuld2018circuit,Farhi2018,Farhi2014}. However, for the dynamical models described in this paper, these problems are best framed within the field of optimal control theory, where a number of methods and results already exist on the optimization of pulses and parameters for quantum models~\cite{Magann2020,Yang2020}. Furthermore, advanced methods such as genetic algorithms~\cite{Mortimer2020} and reinforcement-learning~\cite{Niu2019} have recently begun to garner interest within this field. Thus the intersection of quantum optimal control and quantum machine learning seems a fertile avenue of further research, as recently pointed out in~\cite{Magann2020}. Nevertheless, optimization of the parameters of the model is in general expected to be a hard problem, although the difficulty compared to the corresponding optimizations currently faced by variational algorithms in gate based quantum computers is believed by the authors to be an open question.\\

Another possible avenue of research towards improving the schemes presented here is to reduce the complexity of operation related to having to turn interactions between the different layers on and off by instead employing autonomous methods similar to those already used within the field of quantum error correction. Using such methods to perform quantum operations in a coherent way can be highly non-trivial, but for networks like the one described above where the latter stages of the network are essentially classical processing, strict coherences should not be needed for the network to operate, thus lowering the bar for autonomous implementations of similar networks. Furthermore, a faithful reproduction of the spiking action of biological neurons necessarily requires non-linearities and non-unitary reset. Thus engineered decoherence would also be an essential resource if more closely reproducing the continuous-time dynamical behaviour of classical spiking neurons is the goal. \\

Finally, tapping into the temporality of the neurons presented above also holds great promise. Indeed, it has already been shown that the temporal behaviour of comparatively simpler networks of spins allows for universal quantum computation~\cite{childs2013universal}. Thus we believe that augmenting the neurons of this paper with less constrained and clock-like dynamics combined with tunable, teachable behaviour and perhaps partial autonomy would be a promiseful route towards a neuromorphic architecture capable of solving complicated and interesting problems within quantum learning. Additionally, this approach will further distinguish the spiking quantum neural networks from conventional gate-based approaches, both through temporality and through increased complexity. Indeed, while all neuron operations presented here can be implemented on gate-based quantum computers at a cost of between two and six 2-qubit gates, it is unlikely that the same will hold true for the operation of larger, coupled, parametrized networks. This expectation mirrors the expected advantage of classical special-purpose hardware for neuromorphic computing---it does not perform computations that a general-purpose processor could not perform, but may do so faster and more efficiently. In a similar manner, the operations performed here could be emulated on either gate-based quantum computers or (universal) annealing architectures, but would require some overhead, for instance in the form of requiring multiple control pulses to emulate the single-pulse evolution of the excitation-counting neuron. Conversely, while we expect universal computation with models similar to those presented here to be possible---assuming either sufficiently large networks, like in~\cite{childs2013universal}, or sufficiently reconfigurable couplings---we do not expect such a construction to be an efficient architecture for arbitrary classes of algorithms. In other words, it seems likely that a spiking quantum neural network will tend to naturally implement different operations, and therefore tackle different classes of problems, compared to annealing-based or gate-based algorithms, thus potentially making it a valuable additional tool for quantum machine learning.\\

\section*{Methods}
\subsection*{Simulations of the Dynamics}
The plots depicted on Fig.~\ref{fig:Exc_Neuron} and Fig.~\ref{fig:Phase_Neuron} were generated by numerically simulating the dynamics of the Hamiltonians \eqref{eq:H_Exc} and \eqref{eq:Phase_Detection}, respectively, using the Python toolbox QuTip~\cite{Johansson2012}. Specifically, the system was initialized in states of the form 
\begin{align}
\left| \Psi_0 \right> &= \left| \text{Input} \right> \left| 0 \right> \nonumber \\
 \left| \text{Input} \right>  &\in \left\{ \left| \Phi^{\pm} \right>, \left| \Psi^{\pm} \right> \right\}  \; , 
\end{align}
where the two kets in the notation specifies the state of the input qubits (qubits 1 and 2, first ket) and the output qubit (qubit 3, second ket) separately. The built-in numerical solver of the QuTiP library was then used to find the trajectories resulting from the time-evolution of each of these states:
\begin{align}
\left| \Psi (t) \right> &= \mathrm{e}^{-\mathrm{i} \frac{H}{\hbar} t} \left| \Psi_0 \right> \; . 
\end{align}
Using these trajectories, the evolution of the quantities depicted in the plot could then be calculated, including the expectation values related to the output qubit:
\begin{align}
\text{Output-X}: \left< \Psi(t) \right| \sigma_3^x \left| \Psi(t) \right>\nonumber\\
 \text{Output-Z}: \left< \Psi(t) \right| \sigma_3^x \left| \Psi(t) \right>
\end{align}
and the overlap between the current state of the inputs and their initial state:
\begin{align}
\text{Input-Fidelity}:& \nonumber\\
 \left< \Psi(t) \right| \left(  \left| \Psi_0 \right> \right.& \left. \left< \Psi_0 \right| + \sigma_3^x \left| \Psi_0 \right> \left< \Psi_0 \right| \sigma_3^x  \right) \left| \Psi(t) \right> \; .
\end{align}
Note that the effect of the second term in this expectation-value is to trace out the dependence on the state of the output. The phase- and Hadamard-gates required for the operation of the neurons were implemented by evolving the system using Hamiltonians of the form:
\begin{align}
H_{\text{Phase}} &= \begin{cases} A \sigma_3^z &  \text{Exc. Neuron} \\
B \sigma_3^z &  \text{Phase Neuron} \end{cases} \nonumber\\
H_{\text{Had}} &= \sqrt{2} B \left( \sigma_3^x + \sigma_3^z \right)
\end{align}
for a sufficient amount of time to implement the operation, i.e.:
\begin{align}
\tau_{\text{Phase}} &= \begin{cases} \frac{\pi}{4 A} &  \text{Exc. Neuron} \\
\frac{\pi}{4 B} &  \text{Phase Neuron} \end{cases} \nonumber \\
\tau_{\text{Had}} &= \frac{\pi}{4 B}\; .
\end{align}
Note that all simulations were performed without the simulation of noise and decoherence.\\
\subsection*{Computation of the average Fidelity}
In order to quantify the performance of the neurons, the simulations of the dynamics were combined with tools based on \cite{Nielsen2002} for calculating the average fidelity of operations, thus allowing the operations implemented by the neurons to be compared to the idealized operations defined in Eq.~\eqref{eq:Number_Detection} and Eq.~\eqref{eq:Phase_Detection}. Each of the fidelities was calculated within the subspace consisting of the 6 states appearing in the corresponding definition, with the following additions made in order to fully specify the desired effects of the neurons within this subspace:
\begin{align}
\left| \Psi^{\pm} \right>\left| 1 \right> \, &\rightarrow \, - \mathrm{i} \left| \Psi^{\pm} \right>\left| 0 \right> & & \text{Exc. Neuron} \nonumber\\
\left| \Phi^{-} \right>\left| 1 \right> \, &\rightarrow \,  \mathrm{i} \, (-1)^{n+1} \left| \Phi^{-} \right>\left| 0 \right> & & \text{Phase Neuron}\nonumber\\
\left| \Psi^{-} \right>\left| 1 \right> \, &\rightarrow \, \mathrm{i} \, (-1)^{n+1}  \left| \Psi^{-} \right>\left| 0 \right> &  & \text{Phase Neuron} \; .
\end{align}
In other words, the average fidelity was computed by comparing the implemented operation $U_{\text{Neur}}$ to this idealized operation $U_{\text{ideal}}$ and then averaging over a uniformly distributed ensemble over the space $\mathcal{D}$ spanned by the 6 states used in the definition of the gate:
\begin{align}
F_{\text{av}} \left(U_{\text{Neur}} \right)  = \int_{\mathcal{D}} d\psi \left|\left< \psi \right| U_{\text{ideal}}^\dagger U_{\text{Neur}} \left| \psi \right>\right|^2 \; .
\end{align}
Note that the 6 states only enters in the ensemble of initial states---the simulations employed the full Hilberspace, and infidelity from leakage out of the 6-state space is fully accounted for by the fidelity-metric. In the specific models presented here, this leakage additionally turned out to be relatively negligible, on the order of $1 \cdot 10^{-4}$.

\section*{Data Availability}
The code and datasets used in the current study are available from the corresponding author upon request.

\section*{Acknowledgments}
L.B.K. and N.T.Z. acknowledge funding from the Carlsberg Foundation and the Danish Council for Independent Research (DFF-FNU). M.D. acknowledges support by the U.S. Department of Energy, Office of Science, Office of Advanced Scientific Computing Research, Quantum Algorithms Teams Program. A.A.-G acknowledges support from the Army Research Office under Award No. W911NF-15-1-0256 and the Vannevar Bush Faculty Fellowship program sponsored by the Basic Research Office of the Assistant Secretary of Defense for Research and Engineering (Award number ONR 00014-16-1-2008). A.A.-G.\ also acknowledges generous support from Anders G.\ Frøseth and from the Canada 150 Research Chair Program.

\section*{Author Contributions}
N.T.Z. and L.B.K. formulated the initial goal of the research, and M.D., P.W. and A.A.-G. subsequently participated in a further refinement of the scope and focus of the research. The bulk of the analytical and numerical work of the study as well as the creation of the initial draft was performed by L.B.K, with further significant contributions to the manuscript by M.D. and P.W. and crucial revisions by A.A.-G. and N.T.Z.

\section*{Competing Interests}
The authors declare no competing interests.

\bibliography{spiking_bibliography}

\begin{thebibliography}{10}
\expandafter\ifx\csname url\endcsname\relax
  \def\url#1{\texttt{#1}}\fi
\expandafter\ifx\csname urlprefix\endcsname\relax\def\urlprefix{URL }\fi
\providecommand{\bibinfo}[2]{#2}
\providecommand{\eprint}[2][]{\url{#2}}

\bibitem{waldrop2016chips}
\bibinfo{author}{Waldrop, M.~M.}
\newblock \bibinfo{title}{The chips are down for {Moore}’s law}.
\newblock \emph{\bibinfo{journal}{Nature}} \textbf{\bibinfo{volume}{530}},
  \bibinfo{pages}{144--147} (\bibinfo{year}{2016}).

\bibitem{gantz2012digital}
\bibinfo{author}{Gantz, J.} \& \bibinfo{author}{Reinsel, D.}
\newblock \bibinfo{title}{The digital universe in 2020: Big data, bigger
  digital shadows, and biggest growth in the far east}.
\newblock \emph{\bibinfo{journal}{IDC iView: IDC Analyze the future}}
  \textbf{\bibinfo{volume}{2007}}, \bibinfo{pages}{1--16}
  (\bibinfo{year}{2012}).

\bibitem{hashem2015rise}
\bibinfo{author}{Hashem, I. A.~T.} \emph{et~al.}
\newblock \bibinfo{title}{The rise of “big data” on cloud computing: Review
  and open research issues}.
\newblock \emph{\bibinfo{journal}{Inf. Syst.}} \textbf{\bibinfo{volume}{47}},
  \bibinfo{pages}{98--115} (\bibinfo{year}{2015}).

\bibitem{krizhevsky2012imagenet}
\bibinfo{author}{Krizhevsky, A.}, \bibinfo{author}{Sutskever, I.} \&
  \bibinfo{author}{Hinton, G.~E.}
\newblock \bibinfo{title}{{ImageNet} classification with deep convolutional
  neural networks}.
\newblock In \emph{\bibinfo{booktitle}{Adv. Neural Inform. Process. Syst. 25}},
  \bibinfo{pages}{1097–1105} (\bibinfo{year}{2012}).

\bibitem{sutskever2014sequence}
\bibinfo{author}{Sutskever, I.}, \bibinfo{author}{Vinyals, O.} \&
  \bibinfo{author}{Le, Q.~V.}
\newblock \bibinfo{title}{Sequence to sequence learning with neural networks}.
\newblock In \emph{\bibinfo{booktitle}{Adv. Neural Inform. Process. Syst. 27}},
  \bibinfo{pages}{3104–3112} (\bibinfo{year}{2014}).

\bibitem{preskill2018quantum}
\bibinfo{author}{Preskill, J.}
\newblock \bibinfo{title}{Quantum computing in the {NISQ} era and beyond}.
\newblock \emph{\bibinfo{journal}{Quantum}} \textbf{\bibinfo{volume}{2}},
  \bibinfo{pages}{79} (\bibinfo{year}{2018}).

\bibitem{biamonte2017quantum}
\bibinfo{author}{Biamonte, J.} \emph{et~al.}
\newblock \bibinfo{title}{Quantum machine learning}.
\newblock \emph{\bibinfo{journal}{Nature}} \textbf{\bibinfo{volume}{549}},
  \bibinfo{pages}{195–202} (\bibinfo{year}{2017}).

\bibitem{dunjko2018machine}
\bibinfo{author}{Dunjko, V.} \& \bibinfo{author}{Briegel, H.~J.}
\newblock \bibinfo{title}{Machine learning \& artificial intelligence in the
  quantum domain: a review of recent progress}.
\newblock \emph{\bibinfo{journal}{Rep. Prog. Phys.}}
  \textbf{\bibinfo{volume}{81}}, \bibinfo{pages}{074001}
  (\bibinfo{year}{2018}).

\bibitem{kapoor2016quantum}
\bibinfo{author}{Kapoor, A.}, \bibinfo{author}{Wiebe, N.} \&
  \bibinfo{author}{Svore, K.}
\newblock \bibinfo{title}{Quantum perceptron models}.
\newblock In \emph{\bibinfo{booktitle}{Adv. Neural Inform. Process. Syst. 29}},
  \bibinfo{pages}{3999--4007} (\bibinfo{year}{2016}).

\bibitem{schuld2018circuit}
\bibinfo{author}{Schuld, M.}, \bibinfo{author}{Bocharov, A.},
  \bibinfo{author}{Svore, K.~M.} \& \bibinfo{author}{Wiebe, N.}
\newblock \bibinfo{title}{Circuit-centric quantum classifiers}.
\newblock \emph{\bibinfo{journal}{Phys. Rev. A}}
  \textbf{\bibinfo{volume}{101}}, \bibinfo{pages}{032308}
  (\bibinfo{year}{2020}).

\bibitem{killoran2018continuous}
\bibinfo{author}{Killoran, N.} \emph{et~al.}
\newblock \bibinfo{title}{Continuous-variable quantum neural networks}.
\newblock \emph{\bibinfo{journal}{Phys. Rev. Research}}
  \textbf{\bibinfo{volume}{1}}, \bibinfo{pages}{033063} (\bibinfo{year}{2019}).

\bibitem{tacchino2019artificial}
\bibinfo{author}{Tacchino, F.}, \bibinfo{author}{Macchiavello, C.},
  \bibinfo{author}{Gerace, D.} \& \bibinfo{author}{Bajoni, D.}
\newblock \bibinfo{title}{An artificial neuron implemented on an actual quantum
  processor}.
\newblock \emph{\bibinfo{journal}{npj Quantum Inf.}}
  \textbf{\bibinfo{volume}{5}}, \bibinfo{pages}{26} (\bibinfo{year}{2019}).

\bibitem{steinbrecher2019quantum}
\bibinfo{author}{Steinbrecher, G.~R.}, \bibinfo{author}{Olson, J.~P.},
  \bibinfo{author}{Englund, D.} \& \bibinfo{author}{Carolan, J.}
\newblock \bibinfo{title}{Quantum optical neural networks}.
\newblock \emph{\bibinfo{journal}{npj Quantum Inf.}}
  \textbf{\bibinfo{volume}{5}}, \bibinfo{pages}{60} (\bibinfo{year}{2019}).

\bibitem{monras2016inductive}
\bibinfo{author}{Monràs, A.}, \bibinfo{author}{Sentís, G.} \&
  \bibinfo{author}{Wittek, P.}
\newblock \bibinfo{title}{Inductive supervised quantum learning}.
\newblock \emph{\bibinfo{journal}{Phys. Rev. Lett.}}
  \textbf{\bibinfo{volume}{118}}, \bibinfo{pages}{190503}
  (\bibinfo{year}{2017}).

\bibitem{albarran2018measurement}
\bibinfo{author}{Albarr{\'a}n-Arriagada, F.}, \bibinfo{author}{Retamal, J.~C.},
  \bibinfo{author}{Solano, E.} \& \bibinfo{author}{Lamata, L.}
\newblock \bibinfo{title}{Measurement-based adaptation protocol with quantum
  reinforcement learning}.
\newblock \emph{\bibinfo{journal}{Phys. Rev. A}} \textbf{\bibinfo{volume}{98}},
  \bibinfo{pages}{042315} (\bibinfo{year}{2018}).

\bibitem{maass1997networks}
\bibinfo{author}{Maass, W.}
\newblock \bibinfo{title}{Networks of spiking neurons: The third generation of
  neural network models}.
\newblock \emph{\bibinfo{journal}{Neural Netw.}} \textbf{\bibinfo{volume}{10}},
  \bibinfo{pages}{1659--1671} (\bibinfo{year}{1997}).

\bibitem{kjaergaard2019superconducting}
\bibinfo{author}{Kjaergaard, M.} \emph{et~al.}
\newblock \bibinfo{title}{Superconducting qubits: Current state of play}.
\newblock \emph{\bibinfo{journal}{Annu. Rev. Condens. Matter Phys.}}
  \textbf{\bibinfo{volume}{11}}, \bibinfo{pages}{369--395}
  (\bibinfo{year}{2020}).

\bibitem{kounalakis2018tuneable}
\bibinfo{author}{Kounalakis, M.}, \bibinfo{author}{Dickel, C.},
  \bibinfo{author}{Bruno, A.}, \bibinfo{author}{Langford, N.~K.} \&
  \bibinfo{author}{Steele, G.~A.}
\newblock \bibinfo{title}{Tuneable hopping and nonlinear cross-{Kerr}
  interactions in a high-coherence superconducting circuit}.
\newblock \emph{\bibinfo{journal}{npj Quantum Inf.}}
  \textbf{\bibinfo{volume}{4}}, \bibinfo{pages}{38} (\bibinfo{year}{2018}).

\bibitem{wallraff2007sideband}
\bibinfo{author}{Wallraff, A.} \emph{et~al.}
\newblock \bibinfo{title}{Sideband transitions and two-tone spectroscopy of a
  superconducting qubit strongly coupled to an on-chip cavity}.
\newblock \emph{\bibinfo{journal}{Phys. Rev. Lett.}}
  \textbf{\bibinfo{volume}{99}}, \bibinfo{pages}{050501}
  (\bibinfo{year}{2007}).

\bibitem{Roy2019}
\bibinfo{author}{Roy, K.}, \bibinfo{author}{Jaiswal, A.} \&
  \bibinfo{author}{Panda, P.}
\newblock \bibinfo{title}{Towards spike-based machine intelligence with
  neuromorphic computing}.
\newblock \emph{\bibinfo{journal}{Nature}} \textbf{\bibinfo{volume}{575}},
  \bibinfo{pages}{607--617} (\bibinfo{year}{2019}).

\bibitem{Merolla2014}
\bibinfo{author}{Merolla, P.~A.} \emph{et~al.}
\newblock \bibinfo{title}{A million spiking-neuron integrated circuit with a
  scalable communication network and interface}.
\newblock \emph{\bibinfo{journal}{Science}} \textbf{\bibinfo{volume}{345}},
  \bibinfo{pages}{668--673} (\bibinfo{year}{2014}).

\bibitem{Davies2018}
\bibinfo{author}{Davies, M.} \emph{et~al.}
\newblock \bibinfo{title}{Loihi: A neuromorphic manycore processor with on-chip
  learning}.
\newblock \emph{\bibinfo{journal}{IEEE Micro}} \textbf{\bibinfo{volume}{38}},
  \bibinfo{pages}{82--99} (\bibinfo{year}{2018}).

\bibitem{Benjamin2014}
\bibinfo{author}{Benjamin, B.~V.} \emph{et~al.}
\newblock \bibinfo{title}{Neurogrid: A mixed-analog-digital multichip system
  for large-scale neural simulations}.
\newblock \emph{\bibinfo{journal}{Proc. IEEE}} \textbf{\bibinfo{volume}{102}},
  \bibinfo{pages}{699--716} (\bibinfo{year}{2014}).

\bibitem{gonzalez2019quantized}
\bibinfo{author}{Gonzalez-Raya, T.}, \bibinfo{author}{Solano, E.} \&
  \bibinfo{author}{Sanz, M.}
\newblock \bibinfo{title}{Quantized three-ion-channel neuron model for neural
  action potentials}.
\newblock \emph{\bibinfo{journal}{Quantum}} \textbf{\bibinfo{volume}{4}},
  \bibinfo{pages}{224} (\bibinfo{year}{2020}).

\bibitem{torrontegui2019unitary}
\bibinfo{author}{Torrontegui, E.} \& \bibinfo{author}{Garc{\'\i}a-Ripoll,
  J.~J.}
\newblock \bibinfo{title}{Unitary quantum perceptron as efficient universal
  approximator}.
\newblock \emph{\bibinfo{journal}{EPL}} \textbf{\bibinfo{volume}{125}},
  \bibinfo{pages}{30004} (\bibinfo{year}{2019}).

\bibitem{supic2017simple}
\bibinfo{author}{\v{S}upi\'c, I.}, \bibinfo{author}{Coladangelo, A.},
  \bibinfo{author}{Augusiak, R.} \& \bibinfo{author}{Ac\'in, A.}
\newblock \bibinfo{title}{A simple approach to self-testing multipartite
  entangled states}.
\newblock \emph{\bibinfo{journal}{New J. Phys.}} \textbf{\bibinfo{volume}{20}},
  \bibinfo{pages}{083041} (\bibinfo{year}{2017}).

\bibitem{sasaki2001quantum}
\bibinfo{author}{Sasaki, M.}, \bibinfo{author}{Carlini, A.} \&
  \bibinfo{author}{Jozsa, R.}
\newblock \bibinfo{title}{Quantum template matching}.
\newblock \emph{\bibinfo{journal}{Phys. Rev. A}} \textbf{\bibinfo{volume}{64}},
  \bibinfo{pages}{022317} (\bibinfo{year}{2001}).

\bibitem{sentis2012quantumlearning}
\bibinfo{author}{Sentís, G.}, \bibinfo{author}{Calsamiglia, J.},
  \bibinfo{author}{Muñoz-Tapia, R.} \& \bibinfo{author}{Bagan, E.}
\newblock \bibinfo{title}{Quantum learning without quantum memory}.
\newblock \emph{\bibinfo{journal}{Sci. Rep.}} \textbf{\bibinfo{volume}{2}},
  \bibinfo{pages}{708} (\bibinfo{year}{2012}).

\bibitem{jebara2004probability}
\bibinfo{author}{Jebara, T.}, \bibinfo{author}{Kondor, R.} \&
  \bibinfo{author}{Howard, A.}
\newblock \bibinfo{title}{Probability product kernels}.
\newblock \emph{\bibinfo{journal}{J. Mach. Learn. Res.}}
  \textbf{\bibinfo{volume}{5}}, \bibinfo{pages}{819--844}
  (\bibinfo{year}{2004}).

\bibitem{schuld2019quantum}
\bibinfo{author}{Schuld, M.} \& \bibinfo{author}{Killoran, N.}
\newblock \bibinfo{title}{Quantum machine learning in feature {Hilbert}
  spaces}.
\newblock \emph{\bibinfo{journal}{Phys. Rev. Lett.}}
  \textbf{\bibinfo{volume}{122}}, \bibinfo{pages}{040504}
  (\bibinfo{year}{2019}).

\bibitem{havlicek2019supervised}
\bibinfo{author}{Havl{\'{\i}}{\v{c}}ek, V.} \emph{et~al.}
\newblock \bibinfo{title}{Supervised learning with quantum-enhanced feature
  spaces}.
\newblock \emph{\bibinfo{journal}{Nature}} \textbf{\bibinfo{volume}{567}},
  \bibinfo{pages}{209--212} (\bibinfo{year}{2019}).

\bibitem{Peruzzo2014}
\bibinfo{author}{Peruzzo, A.} \emph{et~al.}
\newblock \bibinfo{title}{A variational eigenvalue solver on a photonic quantum
  processor}.
\newblock \emph{\bibinfo{journal}{Nat. Commun.}} \textbf{\bibinfo{volume}{5}},
  \bibinfo{pages}{4213} (\bibinfo{year}{2014}).

\bibitem{Mcclean2016}
\bibinfo{author}{McClean, J.~R.}, \bibinfo{author}{Romero, J.},
  \bibinfo{author}{Babbush, R.} \& \bibinfo{author}{Aspuru-Guzik, A.}
\newblock \bibinfo{title}{The theory of variational hybrid quantum-classical
  algorithms}.
\newblock \emph{\bibinfo{journal}{New J. Phys.}} \textbf{\bibinfo{volume}{18}},
  \bibinfo{pages}{023023} (\bibinfo{year}{2016}).

\bibitem{Farhi2018}
\bibinfo{author}{Farhi, E.} \& \bibinfo{author}{Neven, H.}
\newblock \bibinfo{title}{Classification with quantum neural networks on near
  term processors.}
\newblock \bibinfo{howpublished}{Preprint at
  \url{https://arxiv.org/abs/1802.06002}} (\bibinfo{year}{2018}).

\bibitem{Farhi2014}
\bibinfo{author}{Farhi, E.}, \bibinfo{author}{Goldstone, J.} \&
  \bibinfo{author}{Gutmann, S.}
\newblock \bibinfo{title}{A quantum approximate optimization algorithm}.
\newblock \bibinfo{howpublished}{Preprint at
  \url{https://arxiv.org/abs/1411.4028}} (\bibinfo{year}{2014}).

\bibitem{Magann2020}
\bibinfo{author}{Magann, A.~B.} \emph{et~al.}
\newblock \bibinfo{title}{From pulses to circuits and back again: A quantum
  optimal control perspective on variational quantum algorithms}.
\newblock \bibinfo{howpublished}{Preprint at
  \url{https://arxiv.org/abs/2009.06702}} (\bibinfo{year}{2020}).

\bibitem{Yang2020}
\bibinfo{author}{Yang, X.-d.} \emph{et~al.}
\newblock \bibinfo{title}{Assessing three closed-loop learning algorithms by
  searching for high-quality quantum control pulses}.
\newblock \bibinfo{howpublished}{Preprint at
  \url{https://arxiv.org/abs/2008.03874}} (\bibinfo{year}{2020}).

\bibitem{Mortimer2020}
\bibinfo{author}{Mortimer, L.}, \bibinfo{author}{Estarellas, M.~P.},
  \bibinfo{author}{Spiller, T.~P.} \& \bibinfo{author}{D'Amico, I.}
\newblock \bibinfo{title}{Evolutionary computation for adaptive quantum device
  design}.
\newblock \bibinfo{howpublished}{Preprint at
  \url{https://arxiv.org/abs/2009.01706}} (\bibinfo{year}{2020}).

\bibitem{Niu2019}
\bibinfo{author}{Niu, M.~Y.}, \bibinfo{author}{Boixo, S.},
  \bibinfo{author}{Smelyanskiy, V.~N.} \& \bibinfo{author}{Neven, H.}
\newblock \bibinfo{title}{Universal quantum control through deep reinforcement
  learning}.
\newblock \emph{\bibinfo{journal}{npj Quantum Inf.}}
  \textbf{\bibinfo{volume}{5}}, \bibinfo{pages}{33} (\bibinfo{year}{2019}).

\bibitem{childs2013universal}
\bibinfo{author}{Childs, A.~M.}, \bibinfo{author}{Gosset, D.} \&
  \bibinfo{author}{Webb, Z.}
\newblock \bibinfo{title}{Universal computation by multiparticle quantum walk}.
\newblock \emph{\bibinfo{journal}{Science}} \textbf{\bibinfo{volume}{339}},
  \bibinfo{pages}{791--794} (\bibinfo{year}{2013}).

\bibitem{Johansson2012}
\bibinfo{author}{Johansson, J.~R.}, \bibinfo{author}{Nation, P.~D.} \&
  \bibinfo{author}{Nori, F.}
\newblock \bibinfo{title}{Qutip: An open-source python framework for the
  dynamics of open quantum systems}.
\newblock \emph{\bibinfo{journal}{Comput. Phys. Commun.}}
  \textbf{\bibinfo{volume}{183}}, \bibinfo{pages}{1760--1772}
  (\bibinfo{year}{2012}).

\bibitem{Nielsen2002}
\bibinfo{author}{Nielsen, M.~A.}
\newblock \bibinfo{title}{A simple formula for the average gate fidelity of a
  quantum dynamical operation}.
\newblock \emph{\bibinfo{journal}{Phys. Lett. A}}
  \textbf{\bibinfo{volume}{303}}, \bibinfo{pages}{249--252}
  (\bibinfo{year}{2002}).

\bibitem{sheldon2016procedure}
\bibinfo{author}{Sheldon, S.}, \bibinfo{author}{Magesan, E.},
  \bibinfo{author}{Chow, J.~M.} \& \bibinfo{author}{Gambetta, J.~M.}
\newblock \bibinfo{title}{Procedure for systematically tuning up cross-talk in
  the cross-resonance gate}.
\newblock \emph{\bibinfo{journal}{Phys. Rev. A}} \textbf{\bibinfo{volume}{93}},
  \bibinfo{pages}{060302} (\bibinfo{year}{2016}).

\end{thebibliography}

\renewcommand{\thesection}{\arabic{section}}
\newpage
\onecolumngrid
\section*{Supplementary Material}
\vspace{0.2cm}

This supplemental material aims to give a more detailed analysis of the spin-models leading to the two neurons presented in the main text, as well as some elaborations on the construction and operation of the Bell-state comparison network. \\

\twocolumngrid

\section{Further details on excitation-parity neuron}
\label{sec:Exc_Neur}

As explained in the main text, the goal with the first neuron is to detect the odd/even parity of the number of excitations (i.e. $\left| \uparrow \right>$-states) in the input. In other words, we wish to be able to distinguish the Bell-states $\left| \Phi^{\pm} \right> = \frac{1}{\sqrt{2}} \left( \left| \uparrow  \, \uparrow  \right> \pm \left| \downarrow \, \downarrow \right>  \right)$ from the states $\left| \Psi^{\pm} \right> = \frac{1}{\sqrt{2}} \left( \left| \downarrow \, \uparrow \right> \pm \left| \uparrow \, \downarrow \right>  \right)$. It turns out that this can be achieved using three ingredients. First, we add a Heisenberg-XXZ interaction between the two input-qubits:
\begin{align*}
H_{\text{input}} &=  \frac{J}{2} \left( \sigma_1^x \sigma_2^x + \sigma_1^y \sigma_2^y  + \gamma \sigma_1^z \sigma_2^z \right) \; ,
\end{align*}
where $J, \beta$ are energies and $\gamma$ is a unitless parameter. The intuition behind this interaction is that it tunes the energy-spectrum of the system in such a way that it allows us to distinguish between the $\left| \Phi^{\pm} \right>$- and $\left| \Psi^{\pm} \right>$-states:
\begin{align*}
H_{\text{input}} \left| \Phi^{\pm} \right> &= \left( \gamma \frac{J}{2} \right)\left| \Phi^{\pm} \right>\\
H_{\text{input}} \left| \Psi^{\pm} \right> &= \left( \pm J -\gamma \frac{J}{2} \right) \left| \Psi^{\pm} \right> \; .
\end{align*}
The next ingredient is to couple the output-qubit to the input-qubits in such a way that the state of the input-qubits influences the energy-spacing of the output-qubit, thus allowing us to do conditional flips of the output qubit by only driving at specific frequencies. Since the property we want to detect has to do with the total $z$-component of the input state (when thought of as a spin state), a reasonable interaction for achieving this would be
\begin{align*}
H_{\text{output}} = \beta \, \sigma_2^z \sigma_3^z \; .
\end{align*}
As a result of this interaction, the eigenstates of the system are no longer the Bell states, but instead take the general form:
\begin{align}
\label{eq:Exc_Spectrum}
\left( A_+ \left| \uparrow \; \downarrow \right> + B_+ \left| \downarrow \; \uparrow \right> \right) \left| \downarrow \right> & & E&=\sqrt{J^2 + \beta^2} - \gamma \frac{J}{2} \nonumber\\
\left( B_+ \left| \uparrow \; \downarrow \right> + A_+ \left| \downarrow \; \uparrow \right> \right) \left| \uparrow \right> & & E&=\sqrt{J^2 + \beta^2} - \gamma \frac{J}{2} \nonumber\\
\left| \downarrow \, \downarrow \right> \left| \downarrow \right>, \left| \uparrow \, \uparrow \right> \left| \uparrow \right>  & & E&= \beta  +  \gamma \frac{J}{2}\\
\left| \downarrow \, \downarrow \right> \left| \uparrow \right>, \left| \uparrow \, \uparrow \right> \left| \downarrow \right>  & & E&= -\beta  +  \gamma \frac{J}{2}\nonumber \\
\left( A_- \left| \uparrow \; \downarrow \right> + B_- \left| \downarrow \; \uparrow \right> \right) \left| \downarrow \right> & & E&=-\sqrt{J^2 + \beta^2} - \gamma \frac{J}{2}\nonumber\\
\left( B_- \left| \uparrow \; \downarrow \right> + A_- \left| \downarrow \; \uparrow \right> \right) \left| \uparrow \right> & & E&=-\sqrt{J^2 + \beta^2} - \gamma \frac{J}{2} \nonumber
\end{align}
for suitable coefficients $A_{\pm}, B_{\pm}$. Assume now that we add as the final ingredient a drive on the output-qubit:
\begin{align}
H_{\text{drive}} &= A \cos\left( \frac{2 \beta}{\hbar} t \right) \sigma_3^x \; .
\label{eq:Exc_Driv}
\end{align}
This drive will resonantly drive the two transitions
\begin{align}
\label{eq:Exc_Transitions}
\left| \uparrow \; \uparrow \right> \left| \downarrow \right> \; \longleftrightarrow \; \left| \uparrow \; \uparrow \right> \left| \uparrow \right> \\
\left| \downarrow \; \downarrow \right> \left| \downarrow \right> \; \longleftrightarrow \; \left| \downarrow \; \downarrow \right> \left| \uparrow \right> \nonumber \; .
\end{align}
On the other hand, all other transitions that this drive could potentially induce will be detuned due to the structure of the energy-spectrum. Specifically, the energy-differences between other pairs of states connected by the $\sigma_3^x$-operator will be
\begin{align*}
\Delta E_0 &= 0\\
\Delta E_\pm &= \pm 2 \sqrt{J^2 + \beta^2} \; ,
\end{align*}
and as a result the driving with frequency $2\beta/\hbar$ will be detuned by
\begin{align*}
\Delta_0 &= \left| 2 \beta \right| \\
\Delta_\pm &= \left| 2 \beta \pm 2 \sqrt{J^2 + \beta^2} \right| \; .
\end{align*}
As long as the strength of the driving is weak compared to these energy-scales, the driving will be unable to induce the corresponding transitions. Thus if we require
\begin{align*}
\left| 2 \beta \right|, \left| 2 \beta \pm 2 \sqrt{J^2 + \beta^2} \right| \gg A\\
\end{align*}
the only effect of the term in \eqref{eq:Exc_Driv} to first order will be to induce the transitions of \eqref{eq:Exc_Transitions}.\\
Let's consider one of these subspaces more closely. To be specific, consider the subspace $\text{Span}\left( \left| \downarrow \; \downarrow \right> \left| \downarrow \right>, \left| \downarrow \; \downarrow \right> \left| \uparrow \right> \right)$. In the basis of these two states, the Hamiltonian reads:
\begin{align*}
H_{\text{subspace}}&= \begin{pmatrix}
\beta + \gamma \frac{J}{2} & A \, \cos\left(\frac{2 \beta}{\hbar} t \right) \\
A \, \cos\left(\frac{2 \beta}{\hbar} t \right) & -\beta + \gamma \frac{J}{2}
\end{pmatrix} \\
&= \begin{pmatrix}
\beta + \gamma \frac{J}{2} & \frac{A}{2} e^{ i \frac{2 \beta}{\hbar} t } + \frac{A}{2} e^{ -i \frac{2 \beta}{\hbar} t }  \\
\frac{A}{2} e^{ i \frac{2 \beta}{\hbar} t } + \frac{A}{2} e^{ -i \frac{2 \beta}{\hbar} t }  & -\beta + \gamma \frac{J}{2}
\end{pmatrix}
\end{align*} 
Performing the unitary transformation
\begin{align*}
U &= \begin{pmatrix}
e^{i \frac{\beta}{\hbar} t} & 0 \\
0 & e^{-i \frac{\beta}{\hbar} t}
\end{pmatrix} e^{i \frac{\gamma J}{2 \hbar} t } \; 
\end{align*}
yields the transformed Hamiltonian
\begin{align*}
H_{\text{trans.}} &= U H U^\dagger + i \hbar \left(\frac{\text{d}}{\text{d}t} U\right) U^\dagger\\
&= \frac{A}{2} \begin{pmatrix}
0 & 1+e^{ i \frac{ 4 \beta}{\hbar} t }\\
1+e^{ -i \frac{ 4 \beta}{\hbar} t } & 0
\end{pmatrix} \; .
\end{align*}
Neglecting the rapidly oscillating terms thus yields an effective Hamiltonian
\begin{align*}
H_{\text{eff.}} &= \frac{A}{2} \, \sigma^x \; .
\end{align*}
Letting this run for a time
\begin{align*}
\tau &= \frac{\pi \hbar}{A}
\end{align*}
yields the transition
\begin{align*}
\left| \downarrow \; \downarrow \right> \left| \downarrow \right> \; \longrightarrow \; -i \left| \downarrow \; \downarrow \right> \left| \uparrow \right> \; ,
\end{align*}
which corresponds to the transition
\begin{align*}
\left| \downarrow \; \downarrow \right> \left| \downarrow \right> \; \longrightarrow \; -i \, e^{i \left( \frac{\beta}{\hbar} - \frac{\gamma J}{2 \hbar} \right) \tau} \left| \downarrow \; \downarrow \right> \left| \uparrow \right> \; ,
\end{align*}
if we undo the unitary transformation.\\
A similar analysis can be performed on the subspace $\text{Span}\left( \left| \uparrow \; \uparrow \right> \left| \downarrow \right>, \left| \uparrow \; \uparrow \right> \left| \uparrow \right> \right)$, and the dynamics of the rest of the states are undisturbed by the driving and thus conforms to the description in eq. \eqref{eq:Exc_Spectrum}. In other words, the effect of waiting the time $\tau$ is that the system performs the operation
\begin{align}
\label{eq:Exc_Phases}
\left| \downarrow \; \downarrow \right> \left| \downarrow \right> \; &\longrightarrow \; -i \, e^{i \frac{1}{\hbar} \left( \beta - \gamma \frac{J}{2 } \right) \tau} \left| \downarrow \; \downarrow \right> \left| \uparrow \right> \nonumber\\
\left| \uparrow \; \uparrow \right> \left| \downarrow \right> \; &\longrightarrow \; -i \, e^{-i \frac{1}{\hbar} \left( \beta + \gamma \frac{J}{2 } \right) \tau} \left| \uparrow \; \uparrow \right> \left| \uparrow \right>\\
\left| \xi_\pm \right> \left| \downarrow \right> \; & \longrightarrow \; e^{i \frac{1}{\hbar} \left( \gamma \frac{J}{2} \pm \sqrt{J^2+\beta^2} \right) \tau } \left|  \xi_\pm \right> \left| \downarrow \right> \; , \nonumber
\end{align}
where $\left| \xi_\pm \right>$ are the states with a single excitation in the input, as sketched in \eqref{eq:Exc_Spectrum}. Note that this already implements an operation akin to the one we want---the output-qubit is flipped if and only if the input-register contains an even number of excitations. However, the fact that different components pick up different phases results in a distortion of the input-states. For instance, the difference in phase between the $\left| \downarrow \; \downarrow \right>$ and $\left| \uparrow \; \uparrow \right>$-states may partially convert a $\left| \Phi^+ \right>$-input to a $\left| \Phi^- \right>$. Additionally, the difference in phase between the states where a flip of the output occurs and those where it does not will distort the relative amplitudes when superpositions are used as input. We would like to pick the parameters in such a way that we avoid these effects, i.e. in such a way that all of the phases are identical. Starting with the two first states, we see that this imposes the restriction
\begin{align*}
& &- i e^{i \frac{1}{\hbar} \left( \beta - \gamma \frac{J}{2 } \right) \tau} &= -i \, e^{-i \frac{1}{\hbar} \left( \beta + \gamma \frac{J}{2 } \right) \tau}\\
\Leftrightarrow & & \; e^{i \frac{1}{\hbar} 2 \beta \tau} &= 1\\
\Leftrightarrow & & \; \frac{1}{\hbar} \, 2 \beta \tau &= 2 \pi k & & \text{ for } k \in \mathbb{Z}\\
\Leftrightarrow & & \; \beta  &=  k A & & \text{ for } k \in \mathbb{Z} \; ,
\end{align*}
Note that this implies
\begin{align*}
e^{i \frac{1}{\hbar} \beta \tau} &= (-1)^k \; .
\end{align*}
Turning to the two other states, requiring identical phases among these implies
\begin{align*}
& & e^{i \frac{1}{\hbar} \left( \gamma \frac{J}{2} + \sqrt{J^2+\beta^2} \right) \tau } &= e^{i \frac{1}{\hbar} \left( \gamma \frac{J}{2} - \sqrt{J^2+\beta^2} \right) \tau }\\
\Leftrightarrow & & \; e^{i \frac{1}{\hbar}  2 \sqrt{J^2+\beta^2}  \tau } &= 1\\
\Leftrightarrow & &  \frac{1}{\hbar} \,  2 \sqrt{J^2+\beta^2}  \tau  &= 2 \pi l & & \text{ for } l \in \mathbb{Z}\\
\Leftrightarrow & &  J &= \pm \sqrt{l^2-k^2} \, A  & & \text{ for } l \in \mathbb{Z}\; .
\end{align*}
Note that in order to assure $J$ is real we are forced to pick $l$ so that it is larger than $k$, and that picking $J$ as prescribed above gives
\begin{align*}
e^{i \frac{1}{\hbar}  \sqrt{J^2+\beta^2}  \tau } = (-1)^l \; .
\end{align*}
At this point, the phases look as follows:
\begin{align*}
\left| \downarrow \; \downarrow \right> \left| \downarrow \right> \; &\longrightarrow \; -i (-1)^k \, e^{- i \frac{1}{\hbar} \gamma \frac{J}{2} \tau} \left| \downarrow \; \downarrow \right> \left| \uparrow \right>\\
\left| \uparrow \; \uparrow \right> \left| \downarrow \right> \; &\longrightarrow \; -i (-1)^k \, e^{- i \frac{1}{\hbar} \gamma \frac{J}{2} \tau} \left| \uparrow \; \uparrow \right> \left| \uparrow \right>\\
\left| \xi_\pm \right> \left| \downarrow \right> \; & \longrightarrow \; (-1)^l e^{i \frac{1}{\hbar} \gamma \frac{J}{2}  \tau } \left|  \xi_\pm \right> \left| \downarrow \right> \; .
\end{align*}
The criterion for identical phases therefore reduce to
\begin{align*}
& & -i (-1)^{k+l} \, e^{- i \frac{1}{\hbar} \gamma J \tau} &= 1\\
\Leftrightarrow & & e^{- i \pi \left( \pm \gamma \sqrt{l^2-k^2}+k+l-\frac{1}{2} \right)} &= 1 \\
\Leftrightarrow & &  \pm \gamma \sqrt{l^2-k^2}-\frac{1}{2}  &=  2 s & & \text{ for } s \in \mathbb{Z} \\
\Leftrightarrow & & \pm \gamma \sqrt{l^2-k^2} &= 2 s - k - l + \frac{1}{2} & & \text{ for } s \in \mathbb{Z} \; ,
\end{align*}
with the sign inherited from the sign of $J$. Note that neither $\gamma=0$ nor $\gamma=1$ allow solutions to this equation-- in both cases, squaring the expression yields an integer on the left-hand side and not on the right-hand side. However, if we allow the factor of $-i$ to be corrected by a subsequent phase-gate, the requirement above in the case $\gamma=1$ reduces to
\begin{align}
\label{Eq:Gamma_One}
\pm \sqrt{l^2-k^2} + (k+l) = 2 s & & \text{ for } s \in \mathbb{Z}
\end{align}
For this to be fulfilled, we will at the very least need $\sqrt{l^2-k^2}$ to be an integer, meaning $k,l$ need to be part of a pythagorean triple. In fact, going through all of the possible combinations of $l$ and $k$ being even or odd yields the fact that whenever $\pm \sqrt{l^2-k^2}$ is an integer it is even (odd) whenever $(k+l)$ is even (odd). In other words, the sum of  these objects is always even when both are integers. Thus it is not just necessary but also sufficient to require $k$ and $l$ to be part of a pythagorean triple for \eqref{Eq:Gamma_One} to be fulfilled. The reason that $\gamma=1$ is especially interesting is that this is required for the operation of the phase-detection neuron (see Sec.~\ref{sec:Phase_Neur} below), and thus using another value of $\gamma$ for our excitation-detection neuron either prohibits the detection of phase or necessitates an implementation where $\gamma$ can be easily tuned.\\
Returning to the general case where $\gamma$ may be arbitrary, we see that the phases match when
\begin{align}
\label{Eq:Gamma_Restriction}
\gamma = \pm \frac{2s-k-l+\frac{1}{2}}{\sqrt{l^2-k^2}} & & \text{ for } s \in \mathbb{Z} \; .
\end{align}
This fully specifies the parameters of the model. However, as alluded to in the case where $\gamma=1$ above, fixing the final phase between the subspace outputting $\left| \downarrow \right>$ and the subspace outputting $\left| \uparrow \right>$ is not essential, since this phase can be adjusted separately through a subsequent phase-gate. Thus \eqref{Eq:Gamma_Restriction} is a less strict requirement than those determining $\beta$ and $J$.

\section{Further details on phase-detection neuron}
\label{sec:Phase_Neur}
The goal of the second neuron is to detect the relative phases of the two components of Bell-states in the computational basis. In other words, we wish to be able to distinguish the states $\left\{ \left| \Phi^+ \right> , \left| \Psi^+ \right> \right\}$ from the states $\left\{ \left| \Phi^- \right> , \left| \Psi^- \right> \right\}$. Similarly to the neuron of the previous section we start by adding a Heisenberg-XXZ interaction among the input-qubits in order to set up an energy-spectrum that distinguishes among the four Bell-states:
\begin{align*}
H_{\text{input}} &=  \frac{J}{2} \left( \sigma_1^x \sigma_2^x + \sigma_1^y \sigma_2^y  + \gamma \sigma_1^z \sigma_2^z \right) \; ,
\end{align*}
yielding the spectrum
\begin{align*}
H_{\text{input}} \left| \Phi^{\pm} \right> &= \left(\gamma \frac{J}{2} \right)\left| \Phi^{\pm} \right>\\
H_{\text{input}} \left| \Psi^{\pm} \right> &= \left(\pm J - \gamma \frac{J}{2} \right)\left| \Psi^{\pm} \right> \; .
\end{align*}
Next, we add an interaction that will tune the energy-spectrum of the output-qubit dependent on the state of the input-qubits. An obvious interaction to use would be one of the form $\sigma_1^x \sigma_2^x \sigma_3^z$, since that exactly tunes the energy it takes to flip the output-qubit in a way that is conditional on the phase encoded in the input. However, as we will see below it is also possible to achieve the same result using only 2-qubit interactions. Specifically, adding a term of the form
\begin{align}
\label{eq:Phase_Input-Output_Coupling}
H_{\text{output}} &= \delta \, \sigma_2^x \sigma_3^x
\end{align}
will also allow us to do the required phase detection. The effect of this term is to couple states with the same phase but different number of excitations. For instance, the state $\left| \Phi^+ \right> \left| + \right>$ becomes coupled to the state $\left| \Psi^+ \right> \left| + \right>$ through a Hamiltonian that in the basis of these two states takes the form
\begin{align*}
H_{\text{eff}} &= \begin{pmatrix}
\gamma \frac{J}{2} & \delta \\
\delta & J - \gamma \frac{J}{2}
\end{pmatrix}
&= \frac{J}{2} \left( \mathbbm{1} + \left( \gamma-1 \right) \sigma^z + \frac{2 \delta}{J} \sigma^x \right)
\end{align*}
Of course a similar coupling of the states $\left| \Phi^+ \right> \left| - \right>$ and $\left| \Psi^+ \right> \left| - \right>$ takes place as well. Indeed, switching the state of the output-qubit is essentially equivalent to switching the sign of $\delta$. Applying similar arguments to the other four states, we arrive at coupling-Hamiltonians with the following structure:
\begin{align}
\label{eq:Phase_Heff}
&\left\{ \left| \Phi^+ \right> \left| \pm \right>, \left| \Psi^+ \right> \left| \pm \right> \right\}: \nonumber \\
&H_{\text{eff}} = \frac{J}{2} \left( \mathbbm{1} + \left( \gamma-1 \right) \sigma^z \pm \frac{2 \delta}{J} \sigma^x \right) \\
&\left\{ \left| \Phi^- \right> \left| \pm \right>, \left| \Psi^- \right> \left| \pm \right> \right\}: \nonumber \\
&H_{\text{eff}} = \frac{J}{2} \left( -\mathbbm{1} + \left( \gamma+1 \right) \sigma^z \pm \frac{2 \delta}{J} \sigma^x \right)  \nonumber
\end{align}
For general $\gamma$, these expressions look relatively symmetric, thus making it hard to fulfil our goal of distinguishing the upper and lower subspaces from each other. The exception to this is whenever $\gamma=\pm 1 $. In this case, the symmetry of the two expressions is very explicitly broken-- one contains a $\sigma^z$-term while the other does not. Let us for definiteness pick $\gamma=1$, resulting in the Hamiltonians
\begin{align*}
H_{\text{eff}} &= \frac{J}{2} \left( \mathbbm{1} \pm \frac{2 \delta}{J} \sigma^x \right) & & \left\{ \left| \Phi^+ \right> \left| \pm \right>, \left| \Psi^+ \right> \left| \pm \right> \right\}\\
H_{\text{eff}} &= \frac{J}{2} \left( -\mathbbm{1} + 2 \sigma^z \pm \frac{2 \delta}{J} \sigma^x \right) & & \left\{ \left| \Phi^- \right> \left| \pm \right>, \left| \Psi^- \right> \left| \pm \right> \right\} \; .
\end{align*}
The eigenstates in the subspaces $\left\{ \left| \Phi^+ \right> \left| \pm \right>, \left| \Psi^+ \right> \left| \pm \right> \right\}$ are then straightforward to write down:
\begin{align}
\label{eq:Phase_Spectrum_1}
\frac{1}{\sqrt{2}} \left( \left| \Phi^+ \right> + \left| \Psi^+ \right> \right) \left| + \right> & & E=\frac{J}{2} + \delta \nonumber\\
\frac{1}{\sqrt{2}} \left( \left| \Phi^+ \right> - \left| \Psi^+ \right> \right) \left| - \right>  & & E=\frac{J}{2} + \delta\\
\frac{1}{\sqrt{2}} \left( \left| \Phi^+ \right> + \left| \Psi^+ \right> \right) \left| - \right> & & E=\frac{J}{2} - \delta \nonumber\\
\frac{1}{\sqrt{2}} \left( \left| \Phi^+ \right> - \left| \Psi^+ \right> \right) \left| + \right>  & & E=\frac{J}{2} - \delta \nonumber
\end{align}
The eigenstates of the subspaces $\left\{ \left| \Phi^- \right> \left| \pm \right>, \left| \Psi^- \right> \left| \pm \right> \right\}$ are in principle more involved. However, in the limit where $J \gg \delta$, the effect of the $\sigma^x$-term in \eqref{eq:Phase_Heff} will be negligible, leading to an effective Hamiltonian of the form
\begin{align*}
H_{\text{eff}} &\simeq \frac{J}{2} \left( -\mathbbm{1} + 2 \sigma^z \right) \; ,
\end{align*}
and thus approximate eigenstates
\begin{align}
\label{eq:Phase_Spectrum_2}
\left| \Phi^- \right> \left| \pm \right> & & E&=\frac{J}{2}\\
\left| \Psi^- \right> \left| \pm \right> & & E&=-\frac{3 J}{2}\; . \nonumber
\end{align}
Note that flipping the state of the output-qubit between $\left| + \right>$ and $\left| - \right>$ does not change the energy of this second batch of states, while flipping the state of the output-qubit changes the energy of the first batch of states \eqref{eq:Phase_Spectrum_1} by the amount $2\delta$. In other words, adding a constant driving-term of the form
\begin{align*}
H_{\text{driv}} &= B \, \sigma_3^z
\end{align*}
to the Hamiltonian will induce resonant flipping of the output-qubit if the input is $\left| \Phi^- \right>$ or $\left| \Psi^-\right>$, while the same driving will be detuned by an amount
\begin{align*}
\Delta =\frac{2 \delta}{\hbar} 
\end{align*}
if the inputs are in the state $\left| \Phi^+ \right>$ or $\left| \Psi^+ \right>$. Assuming this detuning is large compared to the driving-strength $B$ ensures that nothing happens in the latter case, and thus we have exactly what we want: A flipping of the state of the output-qubit if and only if the phases of the Bell-states of the input have a certain value (in this case: $-1$). Assuming we start the output-qubit in the state $\left| + \right>$, evolving the four possible inputs over a time $\tau=\frac{\pi \hbar}{2 B}$ would yield the transitions
\begin{align}
\label{eq:Phase_Neuron_Phases}
&\left| \Phi^- \right>  \left| + \right> \; \longrightarrow \; - i e^{-i \frac{1}{\hbar}\frac{J}{2} \tau } \left| \Phi^- \right> \left| - \right> \nonumber\\
&\left| \Psi^- \right> \left| + \right> \; \longrightarrow \; - i e^{i \frac{1}{\hbar} \frac{3 J}{2} \tau } \left| \Psi^- \right> \left| - \right>\\
&\frac{1}{\sqrt{2}} \left( \left| \Phi^+ \right> \pm \left| \Psi^+ \right> \right) \left| + \right> \nonumber\\
& \hspace*{0.8cm} \longrightarrow \;  e^{-i \frac{1}{\hbar}\left( \frac{J}{2}\pm \delta \right) \tau}  \frac{1}{\sqrt{2}} \left( \left| \Phi^+ \right> \pm \left| \Psi^+ \right> \right) \left| + \right> \, . \nonumber
\end{align}
As described in the section on the excitation-parity neuron, we would like the phases picked up by these terms to match so that the input-amplitudes are not distorted by the evolution of the neuron. Matching the phases on the two first terms yield the criteria
\begin{align*}
& & - i e^{-i \frac{1}{\hbar}\frac{J}{2} \tau } &= - i e^{i \frac{1}{\hbar} \frac{3 J}{2} \tau }\\
\Leftrightarrow & &  e^{i \frac{1}{\hbar} 2J \tau } &= 1\\
\Leftrightarrow & & \frac{1}{\hbar} 2 J \tau &= 2 \pi n & & \text{ for } n \in \mathbb{Z}\\
\Leftrightarrow & & J &= 2 n B & & \text{ for } n \in \mathbb{Z}\; ,
\end{align*}
while  matching the two phases of the last line of \eqref{eq:Phase_Neuron_Phases} yields
\begin{align*}
& & e^{-i \frac{1}{\hbar}\left(\frac{J}{2} - \delta \right) \tau } &= e^{-i \frac{1}{\hbar}\left(\frac{J}{2} + \delta \right) \tau } \\
\Leftrightarrow & &  e^{i \frac{1}{\hbar} 2 \delta \tau } &= 1\\
\Leftrightarrow & &  \frac{1}{\hbar} 2 \delta \tau &= 2 \pi m & & \text{ for } m \in \mathbb{Z}\\
\Leftrightarrow & & \delta &= 2 m B & & \text{ for } m \in \mathbb{Z}\; .
\end{align*}
Note that when these criteria are fulfilled, the following holds:
\begin{align*}
e^{i \frac{1}{\hbar} J \tau} &= \left(-1 \right)^n\\
e^{i \frac{1}{\hbar} \delta \tau} &= \left(-1 \right)^m \; .
\end{align*}
As a result, the phases reduce to
\begin{align*}
&\left| \Phi^- \right> \left| + \right> \; \longrightarrow \; - i e^{-i \frac{1}{\hbar}\frac{J}{2} \tau } \left| \Phi^- \right> \left| - \right> \nonumber\\
&\left| \Psi^- \right> \left| + \right> \; \longrightarrow \; - i e^{i \frac{1}{\hbar} \frac{J}{2} \tau } \left| \Psi^- \right> \left| - \right>\\
&\frac{1}{\sqrt{2}} \left( \left| \Phi^+ \right> \pm \left| \Psi^+ \right> \right) \left| + \right> \nonumber \\
& \hspace*{0.8cm} \longrightarrow \; \left( -1 \right)^m e^{-i \frac{1}{\hbar}\frac{J}{2} \tau} \frac{1}{\sqrt{2}} \left( \left| \Phi^+ \right> \pm \left| \Psi^+ \right> \right) \left| + \right> \; . \nonumber
\end{align*}
Note that we have no parameters left to adjust in order to make these phases identical---indeed, there will be a relative factor $(-i)$ no matter what value of $J$ we use. However, due to our work above the only problematic phases occur between the subspace that flips the output and that which does not. As a result, the phase can be corrected simply by applying a phase-gate on the output-qubit after the operation has finished. Combining this operation with some Hadamards to shift the output-qubit between the computational basis $\left|\downarrow/\uparrow \right>$ and the basis $\left| \pm \right>$ where the flips occur, the full sequence now fulfils our goal of coherently detecting the sign of the Bell-states in the input.\\
Before proceeding, let's briefly review the criteria that need to be fulfilled in order for the neuron to operate in the way explained above. In order for the approximate eigenstates presented in \eqref{eq:Phase_Spectrum_1} and \eqref{eq:Phase_Spectrum_2} to be accurate we need
\begin{align*}
J \left( 1 - \gamma \right) \ll \delta \ll J \left( 1 + \gamma \right) \; .
\end{align*}
Additionally, the detuning-criteria that allows us to drive transitions in only one subspace reduces to
\begin{align*}
B \ll \delta \; .
\end{align*}
Combining this with the phase considerations above yields the requirements
\begin{align*}
\gamma &\simeq 1\\
1 \ll \; &m \, \ll n \; .
\end{align*}
In other words we need the driving to be weak compared to the input-output coupling, which in turn needs to be weak compared to the coupling among the input-qubits, and this strong coupling needs to be of a Heisenberg-XXX type. In practice, it turns out that $m$ in fact does not need to be that large for the scheme to work, probably due to the fact that $\delta$ scales like $2m$ and the detuning scales like $2\delta$. In other words, a more accurate criteria is that
\begin{align*}
1 \ll \; &4m \, \ll 2n \; ,
\end{align*}
which is a much milder criteria on the size of $m$ than the original one. This is supported by the fact that parameters such as $(m,n)=(3,82)$ is able to reach average fidelities of $99.07\%$.\\
It is worth noting that many of the higher-order effects neglected above can have a detrimental effect on the overall fidelity. This is especially true in relation to matching the relative phases of the different input states, since the neglected terms will tend to induce different second-order energy-shifts to the different inputs. As a result, it can at times be fruitful to depart from the criteria described above and tune the interaction-parameters slightly in order to obtain better overall fidelity. For instance, picking  $(m,n)=(2.987,81.99)$ instead of $(3,82)$ increases the fidelity of the operation from $99.07\%$ to $99.59\%$. Even more remarkably, shifting parameters from $(m,n)=(5,80)$ to $(4.985,79.97)$ increases the average fidelity from $96.38\%$ all the way to $99.10\%$, though achieving higher fidelities than this seem require a larger $n$ to match the relatively large $m$.\\
As a final aside, it is worth noting that the form of the input-output interaction presented in \eqref{eq:Phase_Input-Output_Coupling} is far from the oly one that would work. Indeed, the only essential part is that it takes the form
\begin{align*}
\delta \, \sigma_2^x \, \hat{O}_3
\end{align*}
with $\hat{O}_3$ an operator that acts on the third qubit and which breaks the degeneracy and sets up a spectrum with two distinct energy-eigenstates that we can subsequently drive transitions between. For instance, an interaction similar to the cross-resonance interaction favoured by IBM~\cite{sheldon2016procedure}.
\begin{align*}
\delta \, \sigma_2^x \, \sigma_3^z
\end{align*}
would work equally well if paired with driving of the form
\begin{align*}
B \, \sigma^x \; .
\end{align*}
In fact, in this case the eigenstates that we would be inducing flips between would be $\left| \downarrow \right>$ and $\left| \uparrow \right>$ rather than the states $\left| \pm \right>$, which means the Hadamards from the protocol in the main text would no longer be required.

\section{Further details on Bell-state comparison network}
\label{subsec:Final_layer}
The operations of both of the Bell-state comparison networks presented in the main text relied mostly on the application of the two neuron building blocks also presented in the main text. However, both networks required a different operation for the final step of propagating the detected information into the output qubit. This section contains information on how each of these operations can be achieved within the same framework as the neuron models.

\subsection*{Final layer of the main network}
As explained in the main text, most of the Bell-state comparison network in Fig.~\ref{fig:Comparator} of the main text simply consist of iterative applications of the two types of neuron building blocks. However, the final step of the network is lightly different, since it is no longer trying to determine if two bits are equal but rather trying to determine if they are both $\left| \uparrow \right>$, corresponding to the qubits in the third layer having determined both that the phases are equal (blue qubit in $\left|\uparrow\right>$-state) and that the excitation parities are equal (orange qubit in $\left|\uparrow\right>$-state). In other words, the operation of the last layer of the network differs from the excitation-parity neuron in that the output should only be flipped if the inputs are in the state $\left| \uparrow \, \uparrow \right>$ rather than being flipped for both the input $\left| \uparrow \, \uparrow \right>$ and $\left| \downarrow \, \downarrow \right>$\footnote{In the language of gate-based computation, this corresponds to implementing a Toffoli-gate rather than a pair of CNOT's.}. One way to achieve this functionality is to adjust the driving in the excitation-counting neuron. To see how this works, we note first that the driving term used in this neuron can be written as
\begin{align}
\label{eq:Exc_Driving}
 A \cos\left( \frac{2 \beta}{\hbar} t \right) \sigma_3^x =&\, \frac{A}{2} \left( e^{i \frac{2 \beta}{\hbar} t} \sigma_3^+ + e^{-i \frac{2 \beta}{\hbar} t} \sigma_3^- \right) \\
 &+ \frac{A}{2} \left( e^{-i \frac{2 \beta}{\hbar} t} \sigma_3^+ + e^{i \frac{2 \beta}{\hbar} t} \sigma_3^- \right) \; . \nonumber
\end{align}
Looking closely at the arguments in Sec.~\ref{sec:Exc_Neur} reveals that only the first of these terms played a role in driving the transition within the subspace $\text{Span}\left( \left| \downarrow \; \downarrow \right> \left| \downarrow \right>, \left| \downarrow \; \downarrow \right> \left| \uparrow \right> \right)$, while the second term was neglected due to rotating-wave arguments. Similarly, driving the transition within the subspace $\text{Span}\left( \left| \uparrow \; \uparrow \right> \left| \downarrow \right>, \left| \uparrow \; \uparrow \right> \left| \uparrow \right> \right)$ turns out to only involve the second term in \eqref{eq:Exc_Driving}. As a result, using a modified driving of the form
\begin{align}
\label{eq:Final_Layer_Driving}
H_{\text{driv, final}} &= \frac{A}{2} \left( e^{-i \frac{2 \beta}{\hbar} t} \sigma_3^+ + e^{+i \frac{2 \beta}{\hbar} t} \sigma_3^- \right)
\end{align}
would drive only the transition $\left| \uparrow \; \uparrow \right> \left| \downarrow \right> \leftrightarrow \left| \uparrow \; \uparrow \right> \left| \uparrow \right>$, and thus we would only detect the $\left| \uparrow \, \uparrow \right>$-state. The general intuition behind this is that an operator $\hat{O}$ that changes the (unperturbed) energy of the system by $\Delta E$ needs to enter in the Hamiltonian as
\begin{align*}
\Delta H = \hat{O} \; e^{- i \frac{\Delta E}{\hbar} t} + \text{h.c.}
\end{align*} 
in order for the combined term to resonantly drive the transitions that the operator $\hat{O}$ induce. Thus because a $\sigma_3^+$ induce transitions costing $\Delta E = -2\beta$ in the $\left| \downarrow \, \downarrow \right> \left|\downarrow/\uparrow\right>$-subspace and $\Delta E = 2\beta$ in the $\left| \uparrow \, \uparrow \right>\left|\downarrow/\uparrow\right>$-subspace we can easily use this rule of thumb to identify which term drives which transitions. \\
At first, the new reduced driving of \eqref{eq:Final_Layer_Driving} may look more complicated than the original form in \eqref{eq:Exc_Driving}. However, it is possible to extract this type of driving from a term very similar to the one in \eqref{eq:Exc_Driving} using a term corresponding to a local magnetic field on the output-qubit:
\begin{align}
\label{eq:Local_Field}
\frac{\Omega}{2} \sigma_3^z
\end{align}
With this term, flipping the output-qubit from $\left| \downarrow \right>$ to $\left| \uparrow \right>$ now costs the energy $\Omega \pm 2 \beta$, depending on whether the input is in the state $\left| \uparrow \, \uparrow \right>$ or $\left| \downarrow \, \downarrow \right>$. Adding the drive
\begin{align*}
 A \cos\left( \frac{\Omega + 2 \beta}{\hbar} t \right) \sigma_3^x =&\, \frac{A}{2} \left( e^{-i \frac{ \Omega + 2 \beta}{\hbar} t} \sigma_3^+ + e^{i \frac{\Omega + 2 \beta}{\hbar} t} \sigma_3^- \right) \\
 &+ \frac{A}{2} \left( e^{i \frac{\Omega + 2 \beta}{\hbar} t} \sigma_3^+ + e^{-i \frac{\Omega + 2 \beta}{\hbar} t} \sigma_3^- \right) \; , \nonumber
\end{align*}
the first term will then again resonantly induce the transitions $\left| \uparrow \; \uparrow \right> \left| \downarrow \right> \leftrightarrow \left| \uparrow \; \uparrow \right> \left| \uparrow \right>$, as predicted by our rule of thumb. On the other hand, the transitions within the $\left| \downarrow \, \downarrow \right>\left|\uparrow/\downarrow\right>$-subspace will cost $\Delta E = \Omega - 2 \beta$, which does not fit with the driving-frequency of the second term. The conclusion that we can draw from our rule of thumb is in other words that the second term is detuned by an amount
\begin{align*}
\Delta_{\text{FL}} = \frac{1}{\hbar} \left| -\left(\Omega + 2 \beta \right) - \left(\Omega - 2 \beta\right) \right| = \frac{2 \left|\Omega\right|}{\hbar}
\end{align*}
compared to the required value to drive resonant transitions $\left| \downarrow \; \downarrow \right> \left| \downarrow \right> \leftrightarrow \left| \downarrow \; \downarrow \right> \left| \uparrow \right>$. Picking $\Omega$ sufficiently large compared to the strength of the driving should therefore suppress the effect of the second term, leading to effective dynamics of the type we desire.\\
Note that in some platforms adding a local magnetic field of the type \eqref{eq:Local_Field} would not increase the complexity of the protocol, since such energy-shifts between the computational states would be naturally present within the architecture. Indeed, such terms appear naturally in many implementations of superconducting qubits~\cite{kjaergaard2019superconducting}, and would need to be compensated by changing to a rotating picture in order to implement Hamiltonians of the form presented in the main text.\\

As with the two main types of neurons, it would be beneficial if the relative phases related to different inputs could be brought to match. Performing an analysis similar to the one in Sec.~\ref{sec:Exc_Neur} yields that running the interactions above for the time $\tau = \frac{\pi}{A}$ yields the time-evolution
\begin{align}
\left| \downarrow \, \downarrow \right>\left| \downarrow \right> \; &\longrightarrow \; e^{-i \frac{1}{\hbar} \left( \beta + \gamma \frac{J}{2} \right) \tau} \left|\downarrow \, \downarrow \right> \left| \downarrow \right> \nonumber \\
\left| \uparrow \, \uparrow \right> \left| \downarrow \right> \; &\longrightarrow \; -i e^{-i \frac{1}{\hbar} \left( \beta + \gamma \frac{J}{2} \right) \tau} \left| \uparrow \, \uparrow \right> \left| \uparrow \right> \label{eq:Final_Layer_Evo}\\
\left| \xi_{\pm} \right> \left| \downarrow \right>  \; &\longrightarrow \; e^{i \frac{1}{\hbar} \left( \gamma \frac{J}{2} \pm \sqrt{J^2+\beta^2} \right) \tau}  \left| \xi_{\pm} \right> \left| \downarrow \right> \nonumber
\end{align}
where $\left| \xi_{\pm} \right>$ are the one-excitation states also used in Sec.~\ref{sec:Exc_Neur}. Note that making the phases of the first two states match through clever selection of parameters is not possible-- whatever we do, we cannot get rid of the factor $-i$ that appear on the $\left| \uparrow \, \uparrow \right>$-state. Luckily, the usual trick of correcting this phase through a subsequent phase-gate on the output-qubit works just as well here as it did in Sec.~\ref{sec:Exc_Neur} and \ref{sec:Phase_Neur}. Adding this correction, matching the phases reduce to the requirements
\begin{align}
\label{eq:Final_Neur_Phase_Req}
e^{-i \frac{1}{\hbar} \left( \beta + \gamma \frac{J}{2} \right) \tau} &= e^{i \frac{1}{\hbar} \left( \gamma \frac{J}{2} + \sqrt{J^2+\beta^2} \right) \tau}  \\
&= e^{i \frac{1}{\hbar} \left( \gamma \frac{J}{2} - \sqrt{J^2+\beta^2} \right) \tau}  \nonumber \; .
\end{align}
Note that these phases are identical to some of the phases that needed to be matched when dealing with the excitation-parity neuron (see \eqref{eq:Exc_Phases} ) when a phase-gate was applied to the output qubit of that system. From the discussion of that system we therefore already know a set of solutions:
\begin{align}
\label{eq:Solution_subset}
\gamma &= 1 \nonumber \\
\beta &= k A\\
J &= \pm \sqrt{l^2-k^2} \, A \nonumber
\end{align}
where $l$ is the largest number in a Pythagorean triple that also contains $k$. However, looking closer at how this solution came about, we can note two things: Firstly, the problem of matching the phases in Sec.~\eqref{sec:Exc_Neur} contained an additional constraint compared to our current problem. Indeed, the fact that $\beta/A$ should be an integer originated from matching the phases of $\left| \downarrow \; \downarrow \right>$ and $\left| \uparrow \; \uparrow \right>$, a set of phases that in this case are automatically identical once the factor of $-i$ is taken care of. Secondly, we arrived at the solution \eqref{eq:Solution_subset} by assuming $\gamma=1$, a requirement that was only necessary if we also wanted to operate with a phase-neuron on the same input-state. Both of these observations indicate that more general solutions than the one in \eqref{eq:Solution_subset} should exist to our phase-matching problem. Finding these more general solutions from \eqref{eq:Final_Neur_Phase_Req} is relatively straightforward. From the matching of the phases on the $\left| \xi_{\pm} \right>$-states we get
\begin{align*}
& & e^{i \frac{1}{\hbar} \left( \gamma \frac{J}{2} + \sqrt{J^2+\beta^2} \right) \tau}  &= e^{i \frac{1}{\hbar} \left( \gamma \frac{J}{2} - \sqrt{J^2+\beta^2} \right) \tau} \\
\Leftrightarrow & &  e^{i \frac{1}{\hbar} 2 \sqrt{J^2+\beta^2} \tau} &= 1\\
\Leftrightarrow & & \frac{1}{\hbar} 2 \sqrt{J^2+\beta^2}  \tau &=  2 \pi l & & \text{ for } l \in \mathbb{Z}\\
\Leftrightarrow & & J &=  \pm \sqrt{l^2 A^2-\beta^2} & & \text{ for } l \in \mathbb{Z} \; .
\end{align*}
Using the fact that this implies $e^{i \frac{1}{\hbar} 2 \sqrt{J^2+\beta^2} \tau} = (-1)^l$ now yields the final criteria
\begin{align}
\label{eq:Beta_Crit}
& & e^{-i \frac{1}{\hbar} \left( \beta + \gamma \frac{J}{2} \right) \tau} &= \left(-1 \right)^l e^{i \frac{1}{\hbar} \left( \gamma \frac{J}{2} \right) \tau} \\
\Leftrightarrow & & \pm \gamma \sqrt{l^2 A^2 - \beta^2} &= \left( 2 s - l \right) A - \beta & & \text{ for } s \in \mathbb{Z} \; , \nonumber
\end{align}
where the sign is the one appearing in the expression for $J$. We can use this equation to either express $\gamma$ in terms of $\beta$ or express $\beta$ in terms of $\gamma$. The first option is relatively straightforward, yielding simply
\begin{align*}
\gamma &= \pm \frac{\left(2s-l \right) A - \beta}{\sqrt{l^2 A^2 - \beta^2}} \; ,
\end{align*}
analogeously to the expression in \eqref{Eq:Gamma_Restriction}. The second option is a bit more involved. Squaring \eqref{eq:Beta_Crit} yields the quadratic equation
\begin{align}
\label{eq:Beta_Crit_Squared}
0=& \,\left( 1 + \gamma^2 \right) \beta^2 - 2 \left( 2 s - l \right) A \beta \\
&+ \left( \left(2s-l \right)^2 A^2 - l^2 A^2 \gamma^2 \right) \nonumber \; .
\end{align}
This has real solutions whenever
\begin{align}
\label{eq:Beta_Disk}
l^2 \left( 1 + \gamma^2 \right) - \left( 2 s - l \right)^2 \geq 0
\end{align}
in which case they can be expressed as
\begin{align}
\label{eq:Beta_Solutions}
\beta = \, \frac{A}{1+\gamma^2} &\left( \left( 2 s - l \right) \vphantom{sqrt{l^2 \left( 1+ \gamma^2 \right) - \left(2s - l \right)^2}} \right.\\
 &+ \left. \left( -1 \right)^k \gamma \sqrt{l^2 \left( 1+ \gamma^2 \right) - \left(2s - l \right)^2} \right) \nonumber
\end{align}
where $k$ is an integer. Thus because \eqref{eq:Beta_Crit} implies \eqref{eq:Beta_Crit_Squared}, we know that $\beta$ needs to be of this form to fulfil our phase-criteria. However, \eqref{eq:Beta_Crit_Squared} does not in general imply \eqref{eq:Beta_Crit}, and thus it is not a priori obvious that all of the solutions in \eqref{eq:Beta_Solutions} will also be solutions to the original equation. Because the right hand side of \eqref{eq:Beta_Crit} is real, a necessary condition is that the left hand side of this expression should also be real, or equivalently:
 \begin{align}
 \label{eq:Beta_squareroot_crit}
 l^2 A^2 - \beta^2 \geq 0 \; .
 \end{align}
Indeed, if this is the case you can move from the squared criterion to the original one by taking the square root. Thus $\beta$ solves the original requirement for some sign of $J$ if and only if it is of the form \eqref{eq:Beta_Solutions} and is smaller in norm than $l A$. Interestingly, this is automatically the case whenever $\beta$ is real. In other words, it turns out that \eqref{eq:Beta_Disk} implies \eqref{eq:Beta_squareroot_crit}. To see this, let $x=s/l$. The solutions then take the form
\begin{align}
\label{eq:Beta_of_x}
\frac{\beta}{Al} &= \frac{1}{1+\gamma^2} \left( \left( 2 x - 1 \right) \vphantom{\sqrt{(2s-l)^2}} \right.\\
 &+ \left. \left(-1 \right)^k \text{sign}\left(l\right) \gamma \sqrt{\left(1+\gamma^2\right) - \left( 2s-l \right)^2} \right) \; . \nonumber
\end{align}
The interval where this is real follows directly from \eqref{eq:Beta_Disk}:
\begin{align*}
\frac{1}{2}- \frac{1}{2}\sqrt{1+\gamma^2} \leq x \leq \frac{1}{2}+ \frac{1}{2}\sqrt{1+\gamma^2}
\end{align*}
Determining the largest absolute value of the function \eqref{eq:Beta_of_x} on this interval is now a question of simple calculus. Taking the derivative and setting it zero yields 
\begin{align*}
1 &= \left(-1 \right)^k \text{sign}\left(l\right) \gamma \frac{2x-1}{\sqrt{\left(1+\gamma^2\right) - \left(2x-1\right)^2}}
\end{align*}
Squaring this yields the fact that the function can only have local extrema when $(2x-1)^2=1$, i.e. when $2x-1 = \pm 1$. At these points, the absolute value of the function reaches
\begin{align*}
\left| \frac{\beta}{Al} \right| &= \frac{1}{1+\gamma^2} \left| \pm 1 + \left(-1 \right)^k \text{sign}\left(l\right) \gamma  \left| \gamma \right| \right|\\
&= \frac{\left| 1 \pm \left(-1 \right)^k \text{sign}\left(l \gamma \right) \gamma^2 \right|}{1 + \gamma^2} 
\end{align*}
 which is at most equal to one. Similarly, determining the value of the function on the boundary of the interval on which it is defined is straightforward and yields $\left(1+ \gamma^2\right)^{-\frac{1}{2}}$, another number that is at most equal to one. Since we know that the largest absolute values that our function takes must be attained either at points with vanishing derivatives or at the boundaries of the interval on which it is defined, we can now conclude that the absolute value of our function never exceeds one---and thus that the requirement \eqref{eq:Beta_squareroot_crit} is fulfilled.

\subsection*{Final layer of the reduced network}
For the final layer of the reduced network in Fig.~\ref{fig:Reduced_Comparator} in the main text, we would like to perform a gate that detects only the input $\left| \downarrow \, \downarrow \right>$. The conceptually simplest way of performing this detection would be to first perform a pair of NOT-gates to map this state to the state $\left| \uparrow \, \uparrow \right>$, then performing the detection of this state in the way detailed in the previous section. Alternatively, running the excitation parity detection from Sec.~\ref{sec:Exc_Neur} followed by the $\left| \uparrow \, \uparrow \right>$-detection of the previous section similarly detects only the state $\left| \downarrow \, \downarrow \right>$. However, for the sake of practical efficiency and conceptual simplicity, an implementation using similar dynamics to the rest of the paper would be preferable. Fortunately, such an implementation can be constructed using arguments that very closely mirror those of the preceding section. Specifically, inspecting \eqref{eq:Exc_Driving} and now keeping only the term
\begin{align}
H_{\text{driv, red}} &= \frac{A}{2} \left( e^{+i \frac{2 \beta}{\hbar} t} \sigma_3^+ + e^{-i \frac{2 \beta}{\hbar} t} \sigma_3^- \right)
\label{eq:H_driv_red_net}
\end{align}
yields the following mapping after a time $\tau = \frac{\pi}{A}$ has elapsed:
\begin{align*}
\left| \downarrow \, \downarrow \right>\left| \downarrow \right> \; &\longrightarrow \; -i e^{-i \frac{1}{\hbar} \left( \beta + \gamma \frac{J}{2} \right) \tau} \left|\downarrow \, \downarrow \right> \left| \uparrow \right> \\
\left| \uparrow \, \uparrow \right> \left| \downarrow \right> \; &\longrightarrow \;  e^{-i \frac{1}{\hbar} \left( \beta + \gamma \frac{J}{2} \right) \tau} \left| \uparrow \, \uparrow \right> \left| \downarrow \right> \\
\left| \xi_{\pm} \right> \left| \downarrow \right>  \; &\longrightarrow \; e^{i \frac{1}{\hbar} \left( \gamma \frac{J}{2} \pm \sqrt{J^2+\beta^2} \right) \tau}  \left| \xi_{\pm} \right> \left| \downarrow \right> \; ,
\end{align*}
in analogy to the mapping of \eqref{eq:Final_Layer_Evo}. Note that the problem of matching the phases of these components is identical to the problem tackled for the detection of $\left| \uparrow \, \uparrow \right>$ above. Furthermore, identical arguments to those contained in the previous section show that the effect of the driving terms in \eqref{eq:H_driv_red_net} can also be achieved through the addition of a term
\begin{align*}
\frac{\Omega}{2} \sigma_3^z
\end{align*}
to the Hamiltonian, combined with the application of a simple cosine-drive of the form
\begin{align*}
 A \cos\left( \frac{\Omega - 2 \beta}{\hbar} t \right) \sigma_3^x \; .
\end{align*}
In other words, it is possible to implement this detection using a scheme that precisely mirrors the detection of the $\left| \uparrow \, \uparrow \right>$ except for the application of a different driving frequency on the output qubit.

\end{document}